\documentclass[twocolumn,tighten,usenames,dvipsnames]{aastex62_aeb}

\usepackage{graphicx}
\usepackage{verbatim}
\usepackage{amssymb}
\usepackage{amsmath}
\usepackage{xspace}

\newcommand\circp[2]{#1^{\!\!\,\circ}\!\!.#2}

\usepackage[normalem]{ulem}
\usepackage{color}

\definecolor{epcol}{rgb}{0.398, 0.0, 0.797}

\def\del#1{{}}

\shorttitle{Missing Gamma-ray Halos}
\shortauthors{Broderick et al.}

\begin{document}

\title{Missing Gamma-ray Halos and the Need for New Physics in the Gamma-ray Sky}

\author[0000-0002-3351-760X]{Avery E. Broderick}
\affiliation{Department of Physics and Astronomy, University of Waterloo, 200 University Avenue West, Waterloo, ON, N2L 3G1, Canada}
\affiliation{Perimeter Institute for Theoretical Physics, 31 Caroline Street North, Waterloo, ON, N2L 2Y5, Canada}
\email{abroderick@perimeterinstitute.ca}

\author[0000-0003-3826-5648]{Paul Tiede}
\affiliation{Department of Physics and Astronomy, University of Waterloo, 200 University Avenue West, Waterloo, ON, N2L 3G1, Canada}
\affiliation{Perimeter Institute for Theoretical Physics, 31 Caroline Street North, Waterloo, ON, N2L 2Y5, Canada}
\email{ptiede@perimeterinstitute.ca}

\author[0000-0002-2137-2837]{Philip Chang}
\affiliation{Department of Physics, University of  Wisconsin-Milwaukee, 3135 North Maryland Avenue, Milwaukee, WI 53211, USA}
\email{chang65@uwm.edu}

\author[0000-0001-8740-0127]{Astrid Lamberts}
\affiliation{Theoretical Astrophysics, Mailcode 350-17, California Institute of Technology, Pasadena, CA 91125, USA}
\email{lamberts@caltech.edu}

\author[0000-0002-7275-3998]{Christoph Pfrommer}
\affiliation{Leibniz-Institut f{\"u}r Astrophysik Potsdam (AIP), An der Sternwarte 16, 14482 Potsdam, Germany}
\email{cpfrommer@aip.de}

\author[0000-0001-8778-7587]{Ewald Puchwein}
\affiliation{Institute of Astronomy and Kavli Institute for Cosmology, University of Cambridge, Madingley Road, Cambridge, CB3 0HA, UK}
\email{puchwein@ast.cam.ac.uk}

\author[0000-0001-9625-5929]{Mohamad Shalaby}
\affiliation{Department of Physics and Astronomy, University of Waterloo, 200 University Avenue West, Waterloo, ON, N2L 3G1, Canada}
\affiliation{Perimeter Institute for Theoretical Physics, 31 Caroline Street North, Waterloo, ON, N2L 2Y5, Canada}
\affiliation{Leibniz-Institut f{\"u}r Astrophysik Potsdam (AIP), An der Sternwarte 16, 14482 Potsdam, Germany}
\affiliation{Department of Astronomy and Astrophysics, University of Chicago, 5640 S Ellis Ave, Chicago, IL 60637, United States}
\email{mshalaby@perimeterinstitute.ca}

\author[0000-0003-4984-4389]{Maria Werhahn}
\affiliation{Leibniz-Institut f{\"u}r Astrophysik Potsdam (AIP), An der Sternwarte 16, 14482 Potsdam, Germany}
\affiliation{Department of Astronomy and Astrophysics, Berlin Institute of Technology, Hardenbergstraße 36, 10623 Berlin, Germany}

\begin{abstract}
  An intergalactic magnetic field stronger than $3\times10^{-13}$~G would explain the lack of a bright, extended degree-scale, GeV-energy inverse Compton component in the gamma-ray spectra of TeV-blazars.  A robustly predicted consequence of the presence of such a field is the existence of degree-scale GeV-energy gamma-ray halos -- gamma-ray bow ties -- about TeV-bright active galactic nuclei, corresponding to more than half of all radio galaxies. However, the emitting regions of these halos are confined to and aligned with the direction of the relativistic jets associated with gamma-ray sources.  Based on the orientation of radio jets,  we align and stack corresponding degree-scale gamma-ray images of isolated Fanaroff-Riley class I and II objects and exclude the existence of these halos at overwhelming confidence, limiting the intergalactic field strength to $<10^{-15}$~G for large-scale fields and progressively larger in the diffusive regime when the correlation length of the field becomes small in comparison to 1~Mpc.  When combined with prior limits on the strength of the intergalactic magnetic field, this excludes a purely magnetic explanation for the absence of halos. Thus, it requires the existence of novel physical processes that preempt the creation of halos, e.g., the presence of beam-plasma instabilities in the intergalactic medium or a drastic cutoff of the very high energy spectrum of these sources.
\end{abstract}


\section{Introduction}

Very high energy gamma rays (VHEGRs, above 100~GeV) emitted from active galactic nuclei (AGNs) annihilate on the intergalactic infrared background after propagating over cosmological distances \citep{Gould+66,Stec-deJa-Sala:92,Ahar_etal:06,Fermi_EBL2012}.  This results in a population of ultrarelativistic electron-positron pairs (with Lorentz factors of $10^6$), streaming primarily through the intergalactic medium (IGM) in cosmic voids \citep{Gould+66}.  The fate of these pairs remains unclear, depending on the competition between nonlinear saturation of virulent plasma beam instabilities and inverse Compton (IC) cooling via the cosmic microwave background \citep{PaperI,Schlickeiser:2013,Chang:2014}.  Should the latter dominate, it will effectively reprocess the original VHEGR emission of AGNs to lower energies, 1-100~GeV, generating an IC halo.  At these energies, the {\em Fermi Space Telescope} provides a high-resolution (68\% inclusion region of the point spread function, PSF, of $\circp{0}{6}$), high-sensitivity map of the entire sky.

The vast majority of observed gamma-ray bright AGNs are blazars, AGNs with jets that are directed at us \citep{3LAC}.  This identification indicates that the gamma-ray emission is strongly beamed toward us \citep{Push_etal:09} and aligned with the underlying AGN jet.  This anisotropy in the VHEGR emission has already been used to argue for lower limits on the strength of a putative intergalactic magnetic field (IGMF) threading cosmic voids. For a handful of known VHEGR sources, the absorbed VHEGR flux and corresponding IC halo flux have been estimated, and compact ($<\circp{0}{6}$), forward-beamed IC halo components are clearly excluded by {\em Fermi} observations at high significance.  One explanation for this disparity is the presence of a strong IGMF, that is, $\gtrsim3\times10^{-16}$~G, that deflects the pairs out of the line of sight prior to their IC emission \citep{Nero-Semi:09,Nero-Vovk:10,Tayl-Vovk-Nero:11,Taka_etal:11,Vovk+12}.  However, a robust prediction of this picture is the presence of extended, degree-scale IC halos about gamma-ray sources, corresponding to the IC emission missing from the line of sight \citep{1994ApJ...423L...5A,2009PhRvD..80b3010E,BTI}.

The IC halo emission itself is typically exceedingly dim and therefore undetectable for an individual source \citep{1994ApJ...423L...5A,2009PhRvD..80b3010E,BTI}.  Many attempts have been made to stack images from known gamma-ray sources and identify extended gamma-ray excesses, though these have met with little success, due, in part, to uncertainties in the PSF \citep{Ando:2010,FLAT-stack:2013}.  Currently, the lack of any significant extension of the gamma-ray emission about known {\em Fermi} blazars has placed a stronger limit of $>3\times10^{-13}$~G assuming that the gamma-ray jet lifetimes exceed 10~Myr \citep{FermiIGMF:18}.  While shorter jet lifetimes can reduce this limit, lifetimes significantly smaller than those inferred from radio observations of blazars would be inconsistent with the large fraction of nearby blazars observed by {\em Fermi} \citep{3LAC}. 

Recently, we have proposed exploiting the expected anisotropy of the IC halos to circumvent systematic uncertainties \citep{BTII}, and we have used this to exclude the presence of a large-scale, uniform IGMF at more than 4$\sigma$ \citep{BTIII}.  This method relies on the bi-lobed anisotropy that results either from the fact that the electrons and positrons produced by the VHEGR annihilation on the intergalactic infrared background are deflected in opposite directions by a uniform IGMF or from the structure of the initial VHEGR jet in combination with a small-scale tangled IGMF \citep{2010ApJ...719L.130N,2015JCAP...09..065L,BTI,BTII,BTIII,2017JCAP...05..005D}.  The structure in the image both increases the surface brightness of the pair halos and distinguishes them from the confounding systematics arising from the instrument response, subthreshold background sources, and diffuse Galactic contributions.

The degree of anisotropy and the gamma-ray flux are strong functions of the jet orientation.  For blazars, AGNs whose jets are directed at us, this presents a weak anisotropy.  In contrast, for oblique jets (i.e., AGNs with jets more than $30^\circ$ off of the line of sight), the expected IC halo structure is striking \citep{BTI}.  Such oblique jet sources are not, however, observed to be gamma-ray bright as their intrinsic gamma-ray emission is beamed away from us, and they are therefore visible primarily via their radio emission.  Therefore, {\em if we can identify oblique gamma-ray jets and properly orient the images,} any excesses due to IC halos would be detectable with high significance.

Here we employ the unified AGN paradigm to identify, align, and stack the oblique counterparts to the gamma-ray-bright blazars observed by {\em Fermi}.
The fact that the parent population of radio-loud objects and misaligned blazar counterparts can be identified is supported by the same clustering properties of {\em Fermi} blazars (BL Lacertae and flat-spectrum radio quasars, or FSRQs) and radio-loud AGNs \citep{Allevato2014}.
To do this, we utilize existing catalogs of radio jet sources identified in the Very Large Array, Faint Images of the Radio Sky at Twenty-centimeters (VLA FIRST) survey \citep{FIRST_CAT}, from which we obtain 20~cm images that show the radio jet orientation.  Given both the locations and orientations of the oblique jets, we then stack the corresponding {\em Fermi} Large Area Telescope (LAT) observations after aligning them.  This procedure was followed separately for Fanaroff-Riley class I and II (FR I and II) objects, presumably corresponding to BL Lacs and FSRQs, respectively.  We then compare these with the anticipated IC halo signals, associated with their respective gamma-ray AGN object classes.

In neither set of comparisons is any evidence for IC halos found.  Based on our simulated stacked IC halos, we are able to exclude their existence by more than 6$\sigma$.  We explore a variety of potential systematic uncertainties and find that none can adequately explain this nondetection.  Therefore, we interpret this in terms of either a novel spectral cutoff between 100~GeV and 1~TeV in gamma-ray-bright AGN (although we provide a number of observational and theoretical arguments, which make this a very unlikely possibility) or as a result of an additional dissipative process that preempts the IC halo formation after the absorption of VHEGRs and the production of the relativistic pair population.

This paper is organized as follows.
In Section~\ref{sec:method}, the method of selecting and aligning sources with radio jets is described.
In Section~\ref{sec:IC_expected}, details of how the expected anisotropic IC halo signal is computed are given.
In Section~\ref{sec:results}, the expected halo signal from our stacking procedure is given together with a discussion of various potential systematic uncertainties.
In Section~\ref{sec:discussion}, we offer an interpretation of the absence of any evidence of the IC halo signal.
We summarize our conclusions in Section~\ref{sec:conclusion}.

\section{Method} \label{sec:method}
\begin{figure}
  \begin{center}
    \includegraphics[width=\columnwidth]{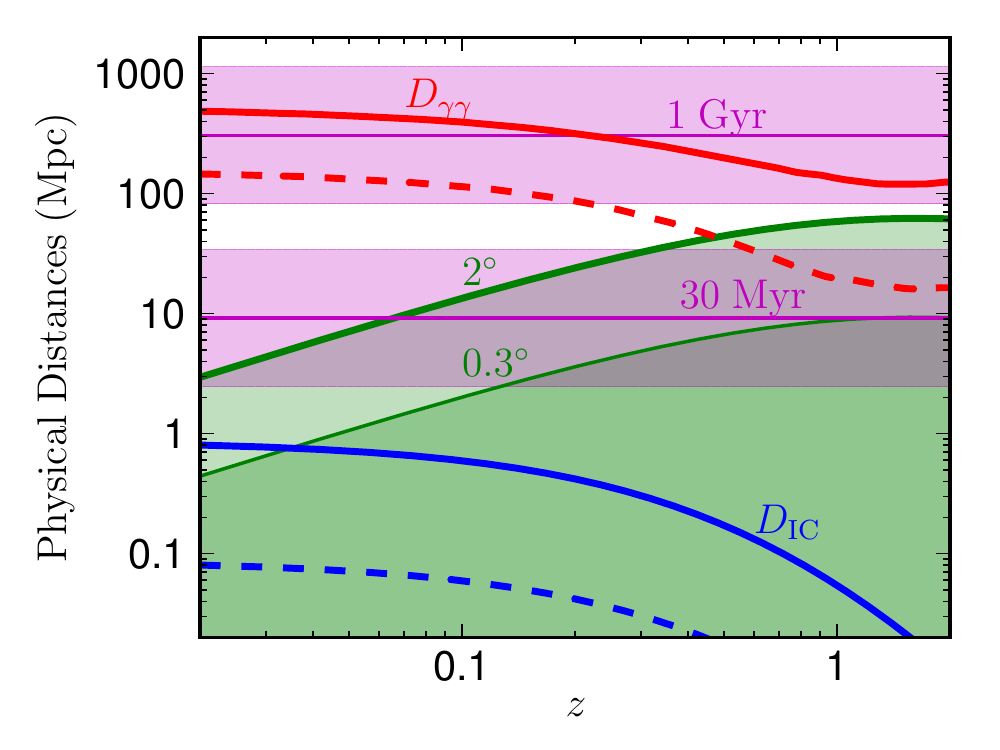}
  \end{center}
  \caption{Characteristic distances that are relevant for the formation and observation of IC halos as a function of redshift.  The physical extent of the $2^\circ$ observing window, to which we restrict our attention in the stacking analysis, is shown by the thick green line; the thin green line indicates the approximate resolution of {\em Fermi} in the 1-100~GeV energy range.  The VHEGR mean free paths implied by \citet{Dominguez11} for the parent gamma rays reponsible for the 1 and 100~GeV IC halo photons (0.9~TeV and 9~TeV, respectively) are shown as solid and dashed red lines, respectively.  The apparent distances to which a VHEGR jet can propagate after 1~Gyr and 30~Myr are shown by the upper and lower purple lines, respectively; the shaded regions indicate the variation with jet inclination between $30^\circ$ and $150^\circ$.  The IC cooling distances of the pairs that produce the 1 and 100~GeV IC halo photons (for electrons and positrons with Lorentz factors of $8\times10^5$ and $8\times10^6$, respectively) are shown by the solid and dashed blue lines, respectively.    
  }\label{fig:distances}
\end{figure}

Typical radio jet/lobe synchrotron lifetimes range from 30 to 100~Myr, placing an empirical lower limit on the timescale over which the jet orientations are stable \citep[see, e.g.,][]{Birzan08}.  This is consistent with a theoretical lower limit, which arises from the difficulty of reorienting the black hole spin, typically requiring the accretion of a mass similar to that of the black hole.  Because the radio mode is associated with accretion rates less than 1\% of the Eddington rate, absent an intervening quasar phase, the jet orientation will be fixed over $\approx100$~Salpeter times (the mass-doubling time scale for an accreting black hole), or 3~Gyr.  Within 1~Gyr, a gamma-ray jet can extend to more than 300~Mpc, or an angular extent larger than $10^\circ$ at $z=1$ (see Figure~\ref{fig:distances}).  Therefore, the orientation of radio jets is expected to be a faithful indicator of the associated IC halo.

Identifying the oblique counterparts of gamma-ray-bright blazars is complicated by the uncertainty in the relationship of blazar properties to their radio morphologies.  Typically, it is assumed that BL Lacs and FSRQs correspond to FR I and II sources, respectively \citep{Padovani+17}.  While there are a number of challenges to this dichotomy \citep{Landt+08,Kharb+10,Kapinska+17}, we assume that it is approximately applicable in what follows.  In practice, this makes little difference: while the BL Lacs are typically intrinsically harder than FSRQs, they are also typically less intrinsically luminous, with the result that the intrinsic luminosity near 1~TeV is similar for both classes.  Nevertheless, we align and stack FR I and II sources separately, comparing them to their canonical blazar counterparts.

Here we describe how radio jet sources are selected (Section~\ref{sec:selection}) and aligned (Section~\ref{sec:align}).

\begin{figure}
  \begin{center}
    \includegraphics[width=\columnwidth]{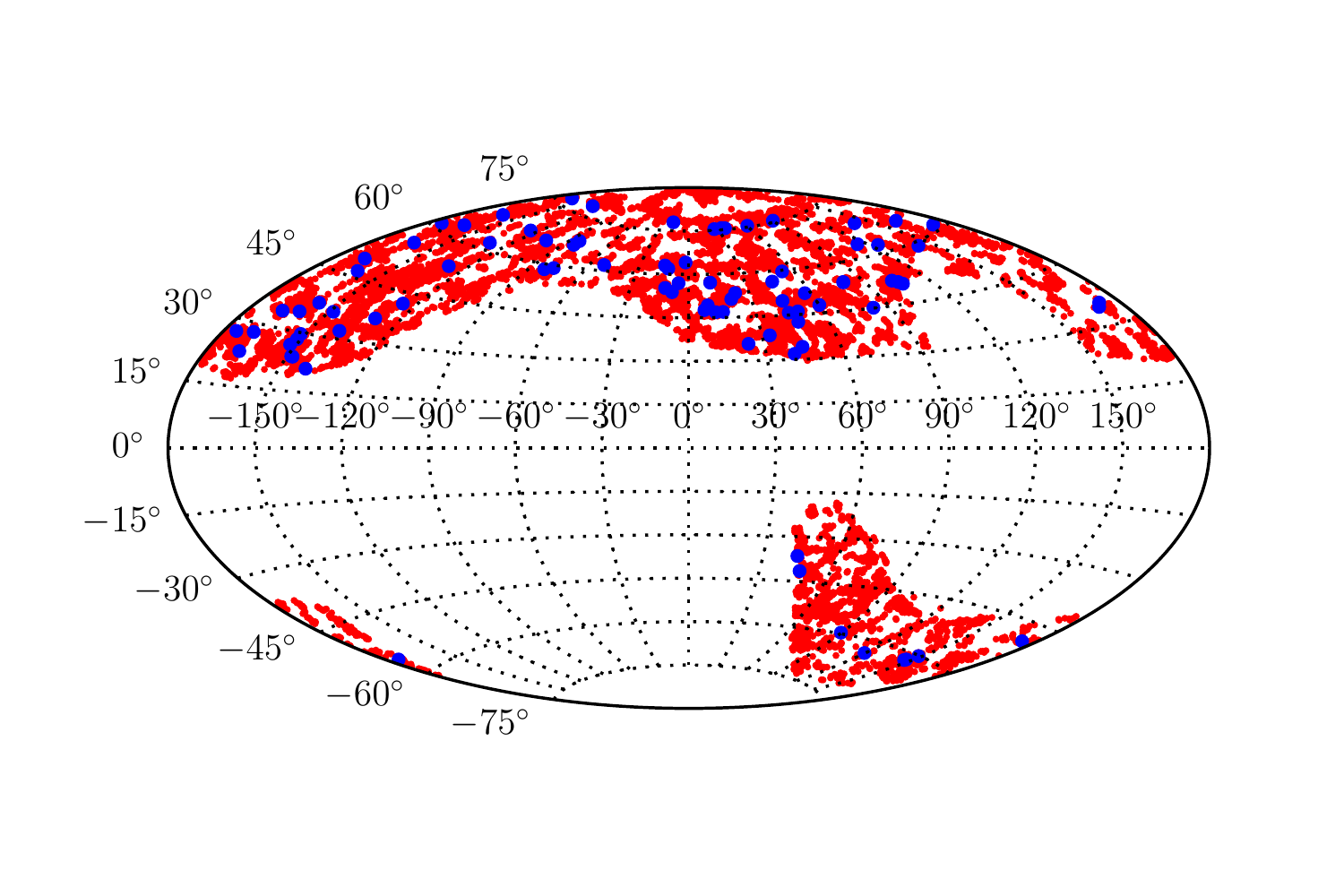}
  \end{center}
  \caption{Aitoff projection of the Galactic coordinates of the FR I (blue) and FR II (red) sources used to generate the stacked analysis.  Both populations are isotropically distributed within the VLA FIRST survey region, which is distributed over a large fraction of the Galactic sky.
  \label{fig:01}}
\end{figure}

\begin{figure*}
  \begin{center}
    \includegraphics[width=\textwidth]{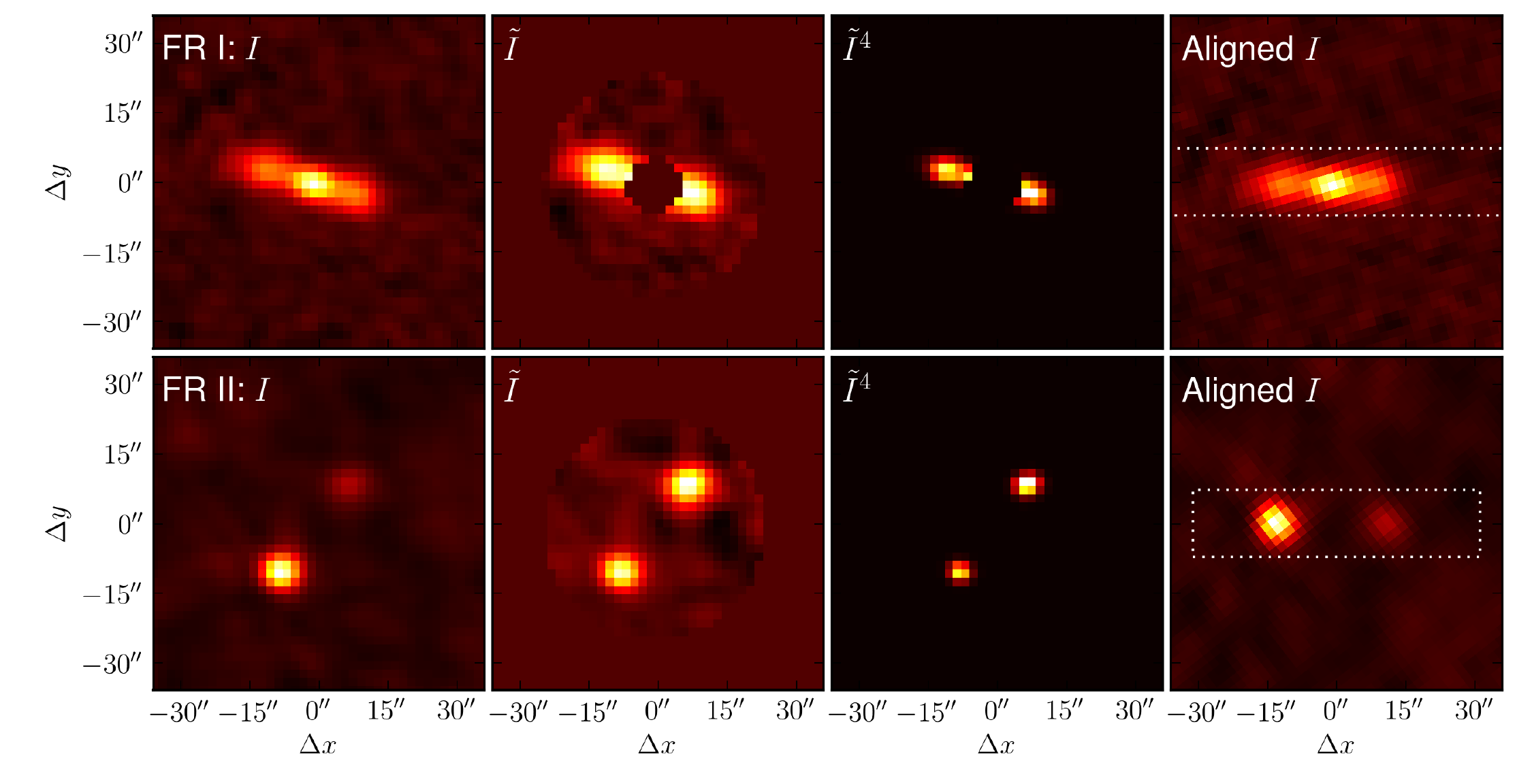}
  \end{center}
  \caption{Stages in the alignment process of the VLA FIRST radio images for an example FR I (top) and FR II (bottom) source.  These include the original radio image after projecting to the pole (far left), after masking and equalizing component fluxes where appropriate (center left), and after dynamically stretching (center right), and the final resulting aligned image (far right).  For reference, in the aligned images, the region to which we require the radio emission to be confined is indicated by the dotted white lines.
  \label{fig:02}}
\end{figure*}

\subsection{Radio Source Selection} \label{sec:selection}

All radio objects we consider are present in the VLA FIRST survey \citep{FIRST_CAT}.  For both classes of radio jets, the procedure is similar: we begin with an existing catalog of jet sources identified in the VLA FIRST fields, from which we exclude radio sources located within $2^\circ$ of existing bright gamma-ray sources (i.e., sources within the {\em Fermi} Large Area Telescope Third Source Catalog, or 3FGL \citealt{3FGL}). This facilitates the detection of a faint halo.  We also exclude objects exhibiting substantial off-axis radio emission, to ensure accurate jet orientation estimates.  For each source remaining, we obtained the most recent averaged VLA FIRST field, flattened the image by projecting it to the pole \citep{BTII}, and determined the jet orientation (see Section~\ref{sec:align}).  We then apply the same transformation to the colocated {\em Fermi} data (Pass 8R2\_V6 ULTRACLEANVETO events within $4^\circ$ of the inferred radio jet origin), which are subsequently aligned and stacked.  A directly comparable set of null results is generated by randomly rotating the {\em Fermi} data.

FR I objects are generally more compact and thus only visible for low redshifts ($z<0.2$), for which we use the FRICAT catalog \citep{FRICAT}, providing 219 sources.  Of these, three are coincident with bright gamma-ray sources, that is, with locations within $\circp{0}{25}$ of a 3FGL member, consistent with that expected from random chance.  After excluding FRICAT sources that have a 3FGL source within $2^\circ$, 87 remain, comprising the sample we employ.  The typical physical extent of the radio jets is 10~kpc, and their near-unity jet/counterjet brightness ratio implies orientations that are significantly oblique.  The number of FR I objects selected in this way is consistent with the number anticipated from the {\em Fermi} Third LAT AGN Catalog \citep[3LAC;][]{3LAC} after correcting for jet opening angle, implying that these represent similar populations.

FR II objects are visible at high redshift ($z>0.1$) via their defining radio lobe emission; thus, radio doubles in the VLA FIRST fields provide a natural catalog of these objects \citep{dubbeltjes}.  Of the 59,192 radio doubles in this catalog, 24,973 meet the internal quality control requirements of the survey  \citep[completeness, angular separations between $18''$ and $1'$, lobe flux ratio;][]{dubbeltjes}. Of these, 356 appear coincident with bright gamma-ray sources, again consistent with random chance.  After excluding sources within $2^\circ$ of existing 3FGL sources, this leaves the 8,741 objects that comprise the set of FR II objects that we employ.  While this is far more numerous than our FR I catalog, due to the much larger redshift ($z\approx2$), the contribution from each source to the stacked IC halo signal is considerably smaller.

\subsection{Aligning Radio Jets} \label{sec:align}

We made use of the most recent (through 2014) averaged images provided by the VLA FIRST Survey\footnote{Available at {\tt http://sundog.stsci.edu/index.html}}.  Within these, we extracted $3'$-radius cutouts around all candidate FR I and FR II sources in the FRICAT and radio double catalog in \citet{dubbeltjes}, with locations shown in Figure~\ref{fig:01}.  These were projected such that the source center locations reported in the relevant catalogs were relocated to the pole, removing the angular aberration associated with equatorial coordinates, as described in Section 5.1 of \cite{BTII} (see Figure~\ref{fig:01}).  The source alignment angle is determined from the radio images such that the jet is oriented along the horizontal axis (see Figure~\ref{fig:02}).  In practice, determining this angle is complicated by the variations in the brightness distribution among components and the radio beam shape.  Here we summarize how this was done; examples of this process for an FR I and FR II are shown in Figure~\ref{fig:02}.

The initial radio images are masked to eliminate unrelated emission outside of $\Delta\theta+\Delta b$, where $\Delta\theta=0.5'$ for FR I objects and is the reported component separation for FR II objects and $\Delta b=0.12'$ is approximately the beam size.  For FR I objects, we additionally mask the central $0.1'$, removing the core.  For the FR II objects, we identify the direction of the brighter component via
\begin{equation}
  \varphi_{\rm CL} = \tan^{-1}\left(\frac{\sum_j I_j\sin\theta_j}{\sum_j I_j\cos\theta_j} \right),
\end{equation}
where $j$ runs over all of the pixels in the masked image, with pixel intensities $I_j$ and $\theta_j$ is the pixel polar angle about the source center.  This is then used to produce an equalized image for FR II objects, for which the radio lobe fluxes can differ by up to an order of magnitude:
\begin{equation}
  \tilde{I}_j = I_j \left[ 1 - \frac{(1-f_F)}{(1+f_F)}\cos(\theta_j-\varphi_{\rm CL}) \right]
\end{equation}
where $f_F$ is the reported component flux ratio.

The extended emission about each component associated with the asymmetric radio beam can significantly bias the estimation of the jet orientation.  Thus, we stretched the dynamic range of the image by finding the orientation of $\tilde{I}^4$.  That is, we estimated the orientation of the double source via
\begin{equation}
  \varphi = \frac{1}{2} \tan^{-1}\left[
    \frac{\sum_j \tilde{I}_j^4 \sin(2\theta_j)}{\sum_j \tilde{I}_j^4 \cos(2\theta_j)}
    \right].
\end{equation}
This was sufficiently accurate that further modeling of the radio beam was unnecessary.

Finally, we performed a second set of quality assessments, removing any sources that
\begin{enumerate}
\item Had any NaN intensity values, indicating that field edges pass through the cropped image region (three FR I and 291 FR II objects).
\item Had any pixels exhibiting a $5\sigma$ or greater flux fluctuation within the region that is more than $0.24'$ off the alignment axis.  These would typically arise from additional components that are not aligned with the primary two (three FR I and 4,624 FR II objects).
\end{enumerate}
This leaves 20,058 FR II objects with reconstructed orientations.  For the subset of these sources without a 3FGL source within $2^\circ$, we performed an additional visual inspection, excluding objects with very nearby off-axis components and dim sources in high-noise regions, removing an additional 128 FR II objects.  Altogether this results in 87 FR I and 8,741 FR II objects that are sufficiently isolated and accurately aligned.

\begin{figure*}
  \begin{center}
    \includegraphics[width=0.32\textwidth]{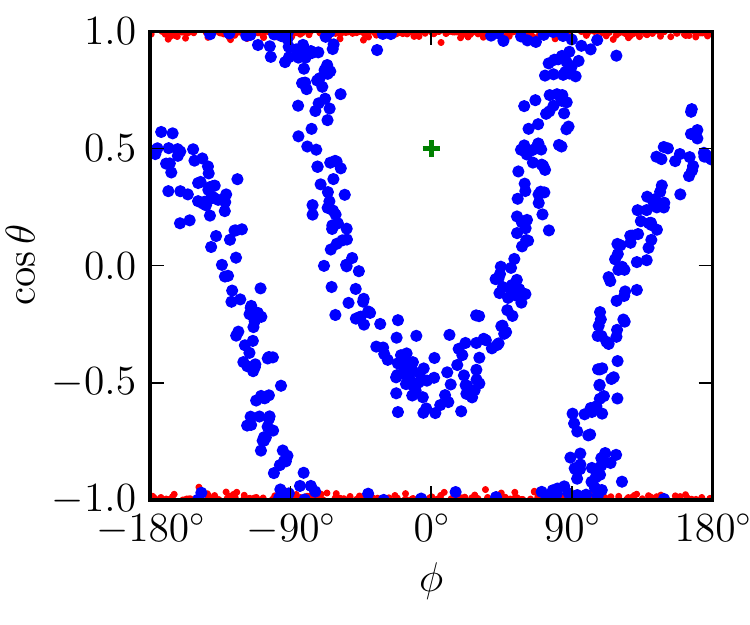}
    \includegraphics[width=0.32\textwidth]{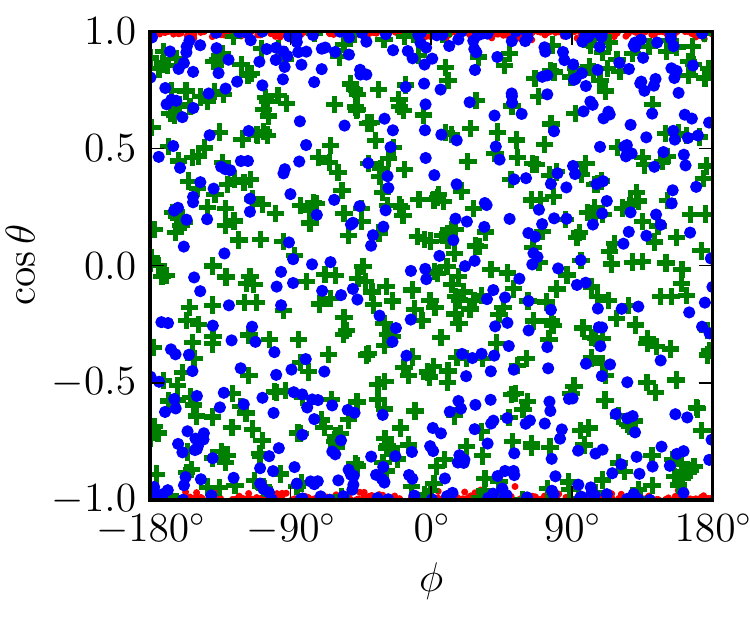}
    \includegraphics[width=0.32\textwidth]{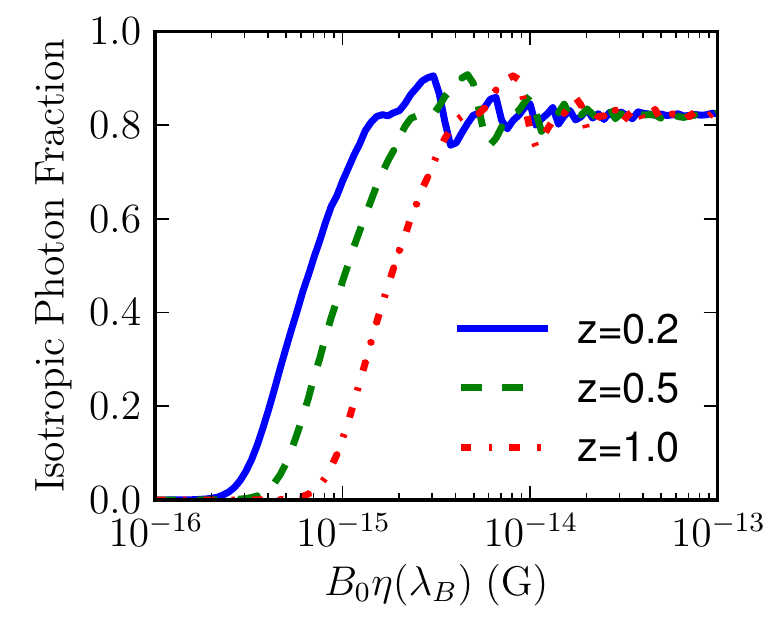}
  \end{center}
  \caption{Left and center: orientations in the unit sphere of IC halo
    photons as seen from the emitting pair source. We show two models,
    corresponding to a fixed (left) and random (center) IGMF.  In both
    plots the initial pair momenta (red points) are normally
    distributed about the azimuthal axis with a standard deviation of
    $5^\circ$.  The IGMF orientations are indicated by the green
    crosses; on the left, a single orientation was chosen, while in the
    center a random orientation was chosen for each IC photon.  As
    such, this figure represents both the gyration of an ensemble
    of photons in a small-scale field and photon gyration about a
    random large-scale field orientation for an ensemble of
    blazars.  The IC halo photons, generated after the pairs gyrate
    through an angle uniformly chosen from
    $(-180^\circ\sin\theta_p,180^\circ\sin\theta_p)$, where $\theta_p$
    is the angle between initial pair momentum and IGMF, are shown by
    the blue points.  In each case, 500 IC photons are drawn.  Right:
    ratio of the number of IC halo photons emitted at more than
    $30^\circ$ from the gamma-ray jet axis to that expected for a
    fully isotropic population as a function of the present-day IGMF
    strength for various redshifts as a function of $B_0 \eta(\lambda_B)$.  When $\lambda_B\gg1~{\rm Mpc}$, this reduces to $B_0$.
    \label{fig:03}}
\end{figure*}

\section{Expected Inverse Compton Halo Signals} \label{sec:IC_expected}
To assess the implications of a nondetection, we estimate the expected IC halo signal independently for the stacked FR I and FR II gamma-ray images. In all cases, the IC emission is confined to the parent VHEGR jet, which is itself aligned with the radio jet. Key inputs include the degree to which we may assume the IC emission is isotropic, the joint redshift and TeV luminosity distributions of the catalog sources, the accuracy of the jet alignment, the local mean free path of VHEGRs within the jet, and the {\em Fermi} PASS 8R2\_V6 ULTRACLEANVETO PSF.  In Section~\ref{sec:Biso}, we demonstrate that we may presume statistical isotropy in the IC emission for IGMF strengths greater than $1\times10^{-15}$~G for our sample, regardless of correlation length.  In Section~\ref{sec:LTeV}, we describe the results of a suite of absorbed \citep[using][]{Dominguez11}, convex, broken-power-law fits to the spectral energy distributions (SEDs) of known gamma-ray-bright BL Lacs and FSRQs, and the resulting redshift-dependent luminosity distributions. The various steps involved in generating an ensemble of simulated stacked halos that are directly comparable with the observed aligned and stacked samples are presented in Section~\ref{sec:ICSim}.

\subsection{Isotropy of Inverse Compton Emission} \label{sec:Biso}
Generally, the IC halos are generated within the VHEGR jet, which is oriented along the radio jet.  However, the direction into which the IC halo gamma rays are emitted varies considerably with the structure and orientation of the IGMF.  Here we describe when it is possible to assume that the IC halo gamma rays are emitted isotropically, that is, emitted equally in all directions, for the stacking analysis we have performed.\footnote{Throughout this section, ``isotropy'' will refer to the distribution of emission directions and should not be confused with the spatial distribution of emission, which is always confined to the VHEGR jet.}

For sufficiently strong and tangled IGMFs, the IC halo is emitted isotropically, a consequence of the large gyration angles about the local IGMF experienced by the parent pair population  \citep{BTI}.  The IC halo generated by a large-scale IGMF is typically emitted highly anisotropically.  For an IGMF that is uniform across the gamma-ray jet, this takes the form of a pair of conical shells, one for each jet, whose orientation and width are set by the IGMF direction and gamma-ray jet half-opening angle, $\approx5^\circ$ \citep{BTII}.  These are shown in the left panel of Figure~\ref{fig:03}, where the gyration angle is randomly chosen uniformly in $(-180^\circ \sin\theta_p, 180^\circ \sin\theta_p)$, in which $\theta_p$ is the initial angle between the IGMF and the pair.  Nevertheless, for our stacking analysis, the assumption that the IC halo gamma rays are emitted isotropically remains statistically true for a large population of jet sources with IGMFs that are randomly oriented relative to the jet axes, shown in the middle panel of Figure~\ref{fig:03} for the same range in gyration angle.  Note that we expect a marginally lower/higher degree of isotropy when considering our FR I/II samples, which have fewer or many more than the 500 objects used for Figure~\ref{fig:03}.

\begin{figure*}
  \begin{center}
    \includegraphics[width=0.45\textwidth]{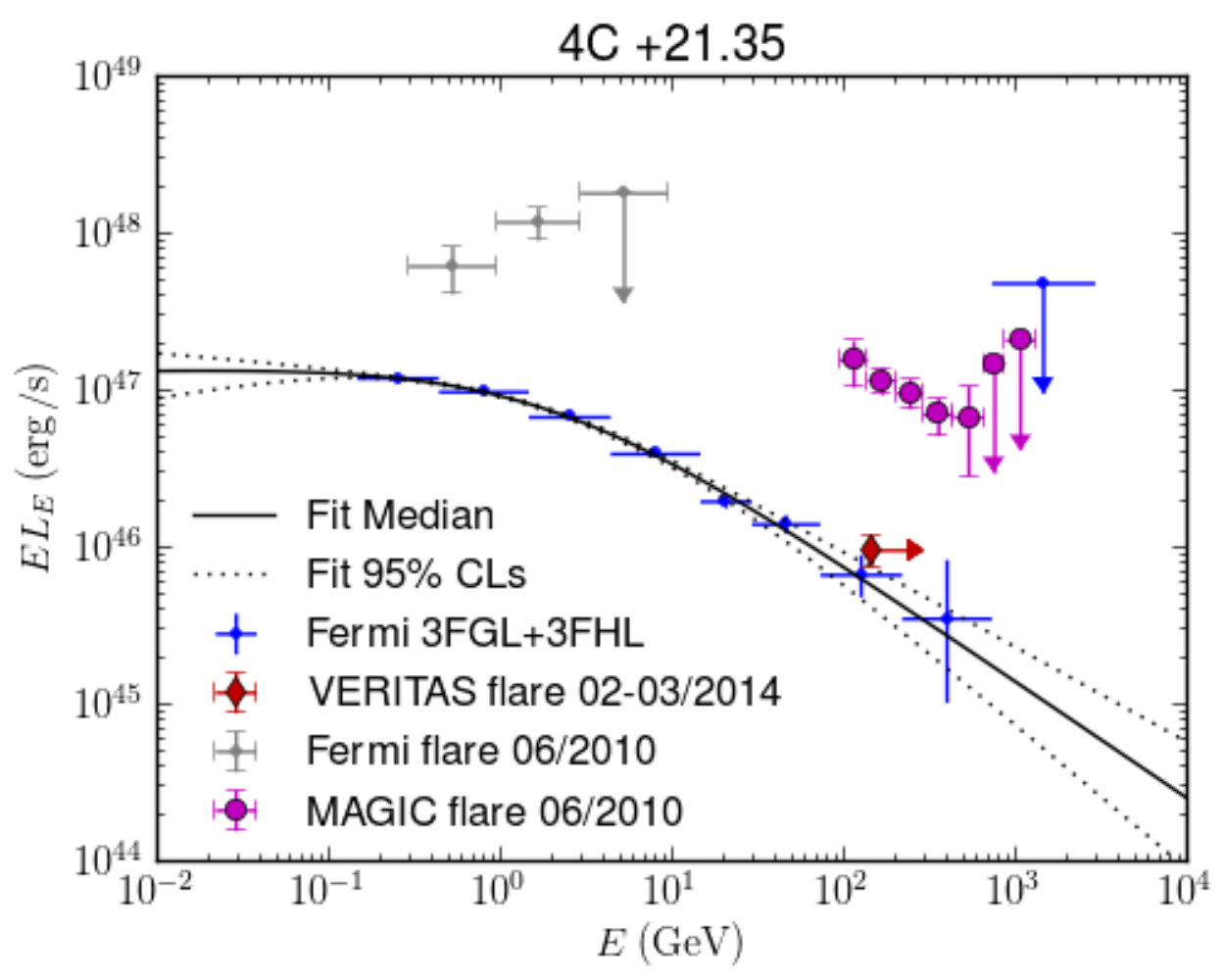}
    \includegraphics[width=0.45\textwidth]{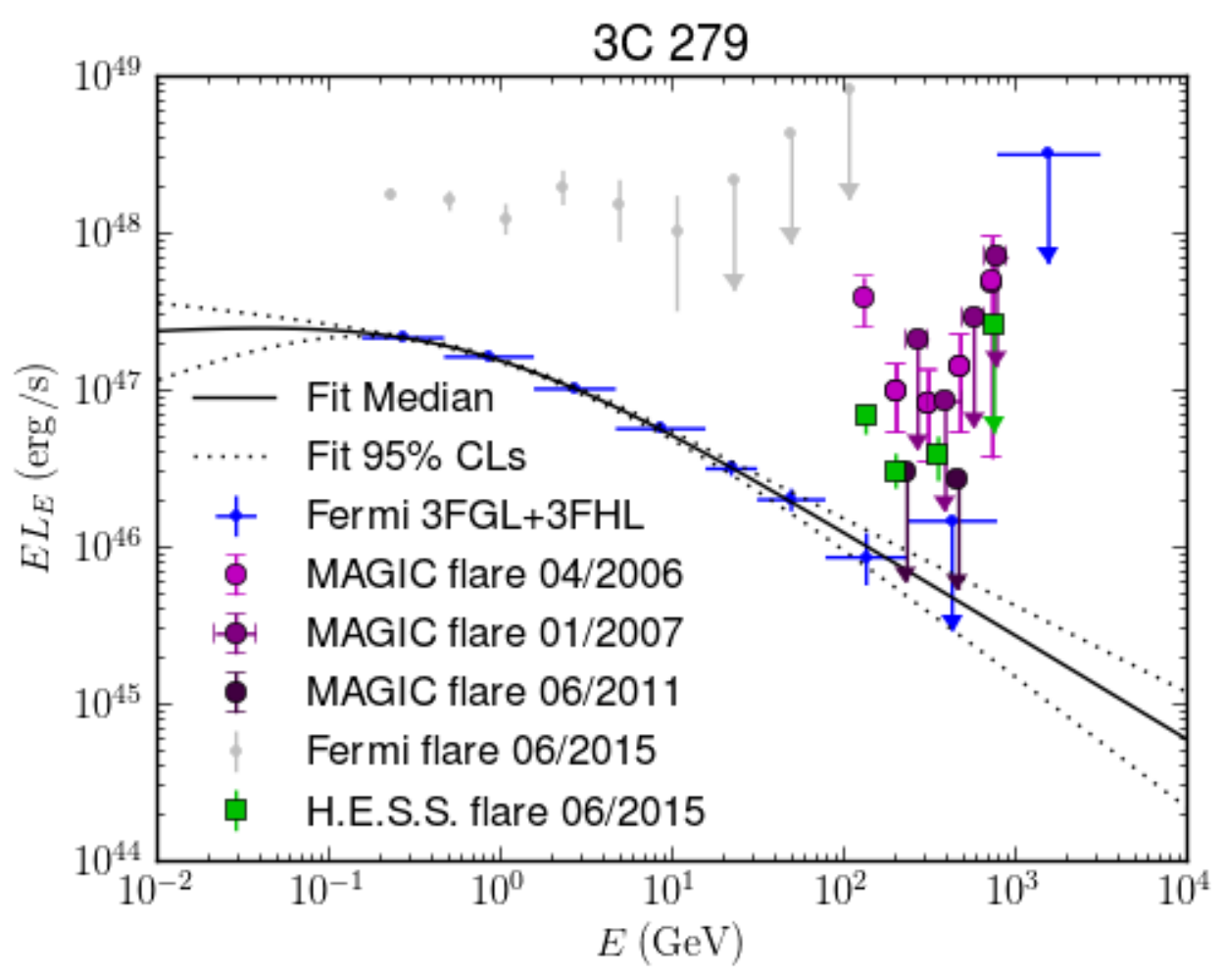}\\
    \includegraphics[width=0.45\textwidth]{fig05c.pdf}
    \includegraphics[width=0.45\textwidth]{fig05d.pdf}
  \end{center}
  \caption{Example SED fits to the combined 3FGL and 3FHL flux data sets shown in the source rest frame and deabsorbed for FSRQs 4C~+21.35 ($z=0.434$) and 3C~279 ($z=0.536$) and BL Lacs Mkn 421 ($z=0.03$) and W Comae ($z=0.102$).  The black lines indicate the median (solid) and one-sided 95\% confidence region (dotted) values at each energy.  The 3FGL (lower four energy bins) and 3FHL (upper five energy bins) are shown in blue; these have been deabsorbed assuming the median SED.  All objects have been detected by air Cerenkov, gamma-ray telescopes during flaring episodes, and the BL Lacs have low-state detections; these are overplotted on the fits.  Note that in all cases the predicted quiescent emission is comfortably less than that observed during flaring epochs and consistent with the low-state fluxes \citep{MAGIC_4c21_2010,VERITAS_4c21_2014,VERITAS_4c21_2014b,MAGIC_3c279_2007,MAGIC_3c279_2011,HESS_3c279_2015,2008ApJ...684L..73A,2009ApJ...707..612A,2015arXiv150807347J,2007ApJ...663..125A}; this is generally true for the sources listed in TeVCat.  For comparison, the TeV luminosity of Mkn 421 is approximately $10^{44}~{\rm erg/s}$.
  \label{fig:04}}
\end{figure*}

The radio sources that we employed are interpreted as radio jets with inclinations greater than $30^\circ$.  Thus, these are predominantly viewed at large angles relative to the jet opening angle. This places lower limits on the gyration angles, and thus the IGMF strength, required to effectively isotropize the IC halo emission.
The typical deflection angle within an IC cooling time for the pairs that generate the 1~GeV IC halo photons is \citep[see][for details]{BTI}
\begin{equation}
  \Delta\alpha_{\rm def}
  \approx
  \frac{\omega_B t_{\rm IC}}{\gamma^2}
  \eta(\lambda_B)
  \approx
  \frac{150^\circ}{(1+z)^2}
  \left(\frac{B_0}{10^{-15}~{\rm G}}\right)
  \eta(\lambda_B)
  \label{eq:alphadefn}
\end{equation}
where $\eta(\lambda_B)$ describes the transition from gyration to diffusion that occurs when the correlation length of the IGMF becomes comparable to the IC cooling time:
\begin{equation}
  \eta(\lambda_B)
  =
  \begin{cases}
    1 & \lambda_B(1+z)^4 \gtrsim 0.9~{\rm Mpc}\\
    \displaystyle \sqrt{\frac{(1+z)^4 \lambda_B}{0.9~{\rm Mpc}}} & \text{otherwise.}
  \end{cases}
  \label{eq:etadefn}
\end{equation}

Note that $\Delta\alpha_{\rm def}$ depends on the redshift both through the dilution of the IGMF and the IC cooling time (through the dilution of the cosmic microwave background), and hence the degree of isotropization will do so as well.  We characterized how isotropized the IC halo emission is (as seen from the source) for a given value of the present-day IGMF strength $B_0$ by the ratio of IC photons at jet viewing angles between $30^\circ$ and $150^\circ$ (consistent with the radio doubles) to the number of IC photons expected from a fully isotropic distribution; this is shown as a function of $B_0$ (assuming $\lambda_B\gg1~{\rm Mpc}$) in Figure~\ref{fig:03}.  By $B_0=10^{-15}\eta^{-1}(\lambda_B)$~G the isotropic fraction reaches 70\% for $z=0.2$, falling to 10\% by $z=1$.  Because it is possible to exclude the presence of IC halos using the FRICAT sample, which includes only sources for which $z<0.2$, we conclude that all present-day IGMF strengths $10^{-15}\eta^{-1}(\lambda_B)$~G are sufficiently isotropized to be constrained.  Above $B_0=10^{-14}\eta^{-1}(\lambda_B)$~G, the isotropic fraction saturates near 80\%.  This falls short of 100\% as a result of the reconcentration of the IC halo photons along the jet axis as they gyrate fully around.  Oscillations arise for the same reason, decreasing in amplitude as the maximum gyration angle becomes many times $\pm360^\circ$.

The angular distributions of the FR I and II objects used in our analysis are roughly uniform within the 10,575 square degrees covered by the VLA FIRST survey (see Figure~\ref{fig:01}).  For FR I objects, the typical angular distance between sources is $11^\circ$, corresponding to an approximate physical distance between sources of 130~Mpc at $z=0.2$.  For FR II objects, the typical angular distance between sources shrinks to $\circp{1}{1}$, but at $z=1$ this corresponds to an approximate physical distance between sources of 200~Mpc.  Thus, even if the current IGMF is ordered on scales up to 100~Mpc, each individual source will see a random realization of the IGMF orientation, effectively isotropizing the IC halo photons in a statistical sense for our 8,741 sources.  Note that while sufficient, IGMF correlation lengths below 100~Mpc are not necessary.  More ordered fields will induce correlations in the IGMF at different source locations, though the typical jet orientations are sufficient to randomize the IC halo photons.  Thus, it is generally true that the resulting IC halo emission will be sufficiently isotropic that the stacked analysis reconstructs the structure of the gamma-ray jets at high fidelity.

\begin{figure*}
  \begin{center}
    \includegraphics[width=0.45\textwidth]{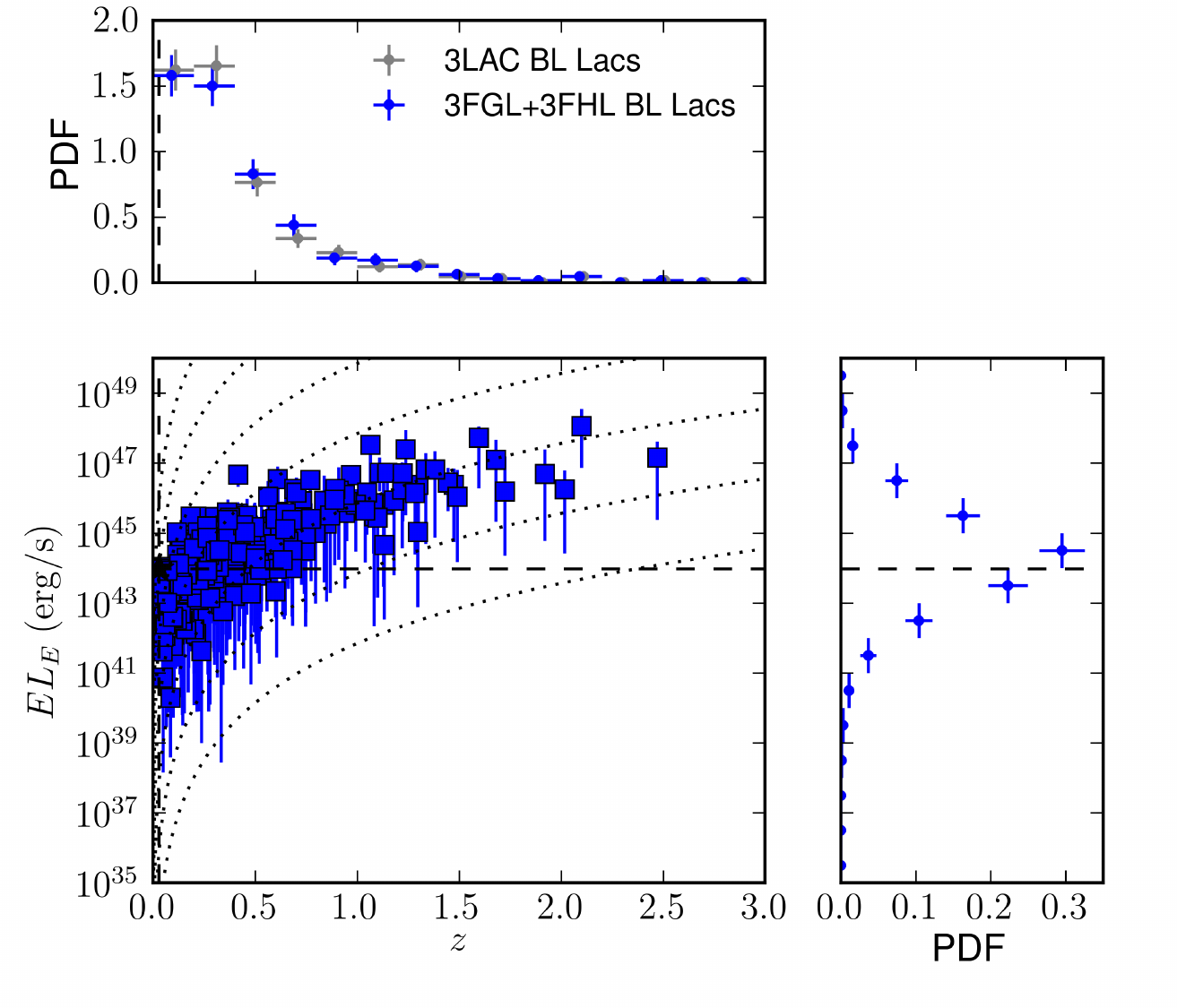}
    \includegraphics[width=0.45\textwidth]{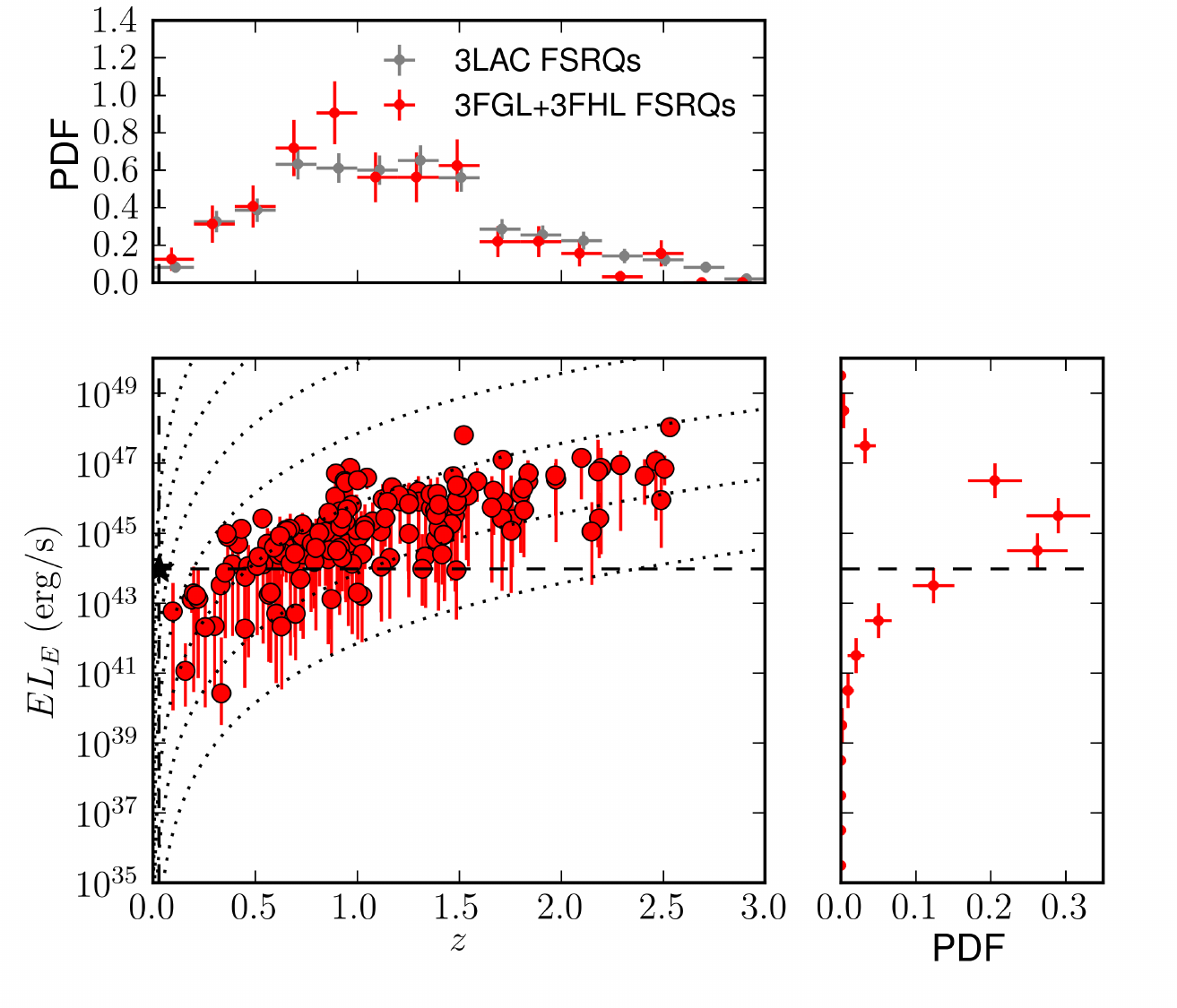}
  \end{center}
  \caption{Joint redshift-TeV luminosity distribution of BL Lac (left) and FSRQ (right) sources present in the 3LAC, 3FGL, and 3FHL catalogs inferred from the convex, broken-power-law fits.  The range for each source indicates the one-sided 95\% confidence level regions.  The projected redshift (top) and luminosity (right) distributions are shown for each.  The former is compared to the distribution of the redshifts of the associated objects within the 3LAC alone (i.e., not confined to those appearing in the 3FHL).  For reference, the position of Mkn 421 is shown (black star, dashed black lines), along with lines of constant flux (dotted black lines).  Note that Mkn 421 is very bright only because of its proximity.
  \label{fig:05}}
\end{figure*}

\subsection{Empirical Joint Redshift-TeV Luminosity Distribution} \label{sec:LTeV}
A key input to simulating the IC halo components is the estimation of the {\em intrinsic}, {\em rest-frame} TeV luminosity ($L_{\rm TeV}$) of gamma-ray-bright AGNs.  This depends on redshift and is complicated by the general precipitous decline in photon flux at high energies.

Estimates of $L_{\rm TeV}$ were obtained by fitting absorbed \citep{Dominguez11}, convex broken-power-law SEDs,
\begin{equation}
  \frac{dN}{dEdt} = \frac{e^{-\tau(z,E)} f}{(E/E_b)^{\Gamma_1} + (E/E_b)^{\Gamma_2}},
  \label{eq:SEDfitfunc}
\end{equation}
to the reported {\em Fermi} fluxes for objects appearing in both the 3LAC and 3FHL with known redshifts.  To do this, we constructed a joint SED from their reported band-specific fluxes in the 3FGL (0.1-0.3~GeV, 0.3-1~GeV, 1-3~GeV, 3-10~GeV)  and 3FHL (10-20~GeV, 20-50~GeV, 50-150~GeV, 150-500~GeV, 0.5-2~TeV).  The absorbed power-law model was integrated across the various gamma-ray bands, and a likelihood was constructed that included the asymmetric nature of the errors and upper limits imposed by nondetections.  This was sampled via the {\it emcee} Markov chain Monte Carlo (MCMC) package \citep{emcee}, generating 160,000 realizations, resulting in converged estimates for $f$, $E_b$, $\Gamma_1$ and $\Gamma_2$.  Examples of these fits are shown in comparison to the deabsorbed {\em Fermi} fluxes, as well as ancillary VHEGR observations, are shown in Figure~\ref{fig:04}, and salient details of the distribution of fitted parameters are presented in Appendix~\ref{app:fits}.  From these, we generate a chain of $\left. E L_E \right|_{E=1~{\rm TeV}} = \left.E^2 dN/dEdt \right|_{E=1~{\rm TeV}}$, discarding the brightest and dimmest 5\% to eliminate outliers.  The resulting luminosity distributions for BL Lac and FSRQ objects are shown in Figure~\ref{fig:05}.

The typical luminosities are comparable to or larger than that of Mkn 421, which is itself extraordinary only in its proximity.  The number of halo photons is proportional to $L_{\rm TeV}$ and is normalized to simulations appropriate for Mkn 421 at $z=0.3$ viewed at an inclination of $60^\circ$ from the jet axis, for which the number of halo photons present within $2^\circ$ of the source is $n_{\rm ref}=16$ photons \citep{BTI}.  This is modified as appropriate for different viewing angles ($\Theta$) and redshifts:
\begin{equation}
  n(\Theta,z) =  \frac{n_{\rm ref}}{1-e^{-\tau_2(60^\circ,0.3)}} \frac{D_L(0.3)^2}{D_L(z)^2} \frac{L_{\rm TeV}}{L_{\rm Mkn 421}},
\end{equation}
where
\begin{equation}
  \tau_2(\Theta,z) = 2^\circ \frac{\pi}{180^\circ} \frac{D_A(z)}{D_{\gamma\gamma}(z)\sin\Theta},
\end{equation}
$D_L(z)$ and $D_A(z)$ are the luminosity and angular diameter distances associated with the WMAP cosmology, and $D_{\gamma\gamma}(z)$ is the absorption mean free path in physical units at redshift $z$, constructed from the observation optical depth in \citet{Dominguez11} (see Appendix \ref{sec:Dpp} for details).  Because we impose limits on the size of the reconstructed halo at a later step in the halo simulation, there is not an additional factor of $1-e^{-\tau_2(\Theta,z)}$ in the above normalization.

All BL Lac objects appear in the 3FHL.  However, only 34\% of FSRQs have detected emission above 10~GeV.  This is expected given the systematically higher redshifts of the FSRQs, corresponding to catastrophic levels of absorption.  Nevertheless, we make the conservative estimate that those not appearing in the 3FHL have $L_{\rm TeV}=0$.

\subsection{Simulating and Stacking Expected IC Halos} \label{sec:ICSim}

We generate simulations of the stacked image in three steps for the front- and back-converted events separately: generate anticipated halo photons, generate observed background photons, and convolve these with the {\em Fermi} Pass 8R2\_V6 ULTRACLEANVETO PSF.  All of these are performed on a sky patch with radius of at least $5^\circ$, which we then crop to the desired regions.  This procedure is performed independently for front- and back-converted events to generate a simulated realization of the expected IC halo signal for each event class, assuming equal sensitivities for both components of the {\em Fermi} Large Area Telescope.

Halo photons are generated by looping over each quality-assured radio jet region, generating a realization for each, and stacking the result.  This is accomplished by
\begin{enumerate}
\item Pulling a random redshift, $z$, from the known distribution (FR I) or from a flux-limited quasar redshift distribution, described in Appendix \ref{sec:FRIIzs} (FR II).
\item Pulling a random jet inclination, $\Theta$, between $30^\circ$ and $150^\circ$, isotropically, consistent with the jet/counterjet brightness ratios in the radio jet catalogs employed.
\item Pulling a TeV luminosity from the joint $z$-$L_{\rm TeV}$ estimated for known BL Lacs and FSRQs in the {\em Fermi} AGN sample.  In doing this, we ensure that the redshift range is sufficiently large to include $\ge10$ example sources and randomly select a TeV luminosity from the individual fit chains, thereby including the underlying fit uncertainty.  At small redshifts, $\Delta z=0.2$, while at large redshifts it can grow to more than 0.5.
\item Pulling the number of expected halo photons, $N$, from a Poisson distribution with mean given by $n(\Theta,z)$.  When including the impact of beam-plasma instabilities, we reduce this number by a factor of $(1+\Gamma_{\rm plasma}/\Gamma_{\rm IC})^{-1}$, where $\Gamma_{\rm plasma}$ and $\Gamma_{\rm IC}$ are the beam-plasma and IC cooling rates, respectively \citep{PaperI}.
\item Pulling $N$ random angular radii from the exponential distribution, $\propto e^{-D_A(z) \theta/D_{\gamma\gamma}(z)\sin\Theta}$, associated with the probability of pair production at a projected angular distance $\theta$ from the source.
\item Pulling $N$ random orientations, $\phi$, from a normal distribution with standard deviation of $5^\circ$, consistent with the estimated accuracy of the alignment and gamma-ray jet width.
\item Constructing the set of photon locations, $(x,y)=(\theta\cos\phi,\theta\sin\phi)$.
\item Computing the propagation time of each candidate photon if a jet lifetime is imposed, $t=[D_A(z)/(c\sin\Theta)]\theta(1\pm\cos\Theta)$, and excluding those above the desired limit.
\item Applying a random shift consistent with the appropriate (front/back) PASS 8R2\_V6 ULTRACLEANVETO PSF averaged over the SED of the gamma-ray background \citep[$\propto E^{-2.38}$;][]{BTII}.
\end{enumerate}

An additional set of background photons are generated such that the total number of photons within $4^\circ$ matches that observed.  The density of the background photons is varied inside and outside $2^\circ$; this is a result of the exclusion of FR I/II objects within $2^\circ$ of 3FGL sources. Again, we apply a random shift consistent with the background-averaged PASS 8R2\_V6 ULTRACLEANVETO PSF \citep{BTII}.  Note that because the number of background photons is constrained to reproduce the total number of gamma rays observed, the radial distribution in the presence of IC halos need not match that found if IC halos are neglected.

\begin{figure*}
  \begin{center}
    \includegraphics[width=\textwidth]{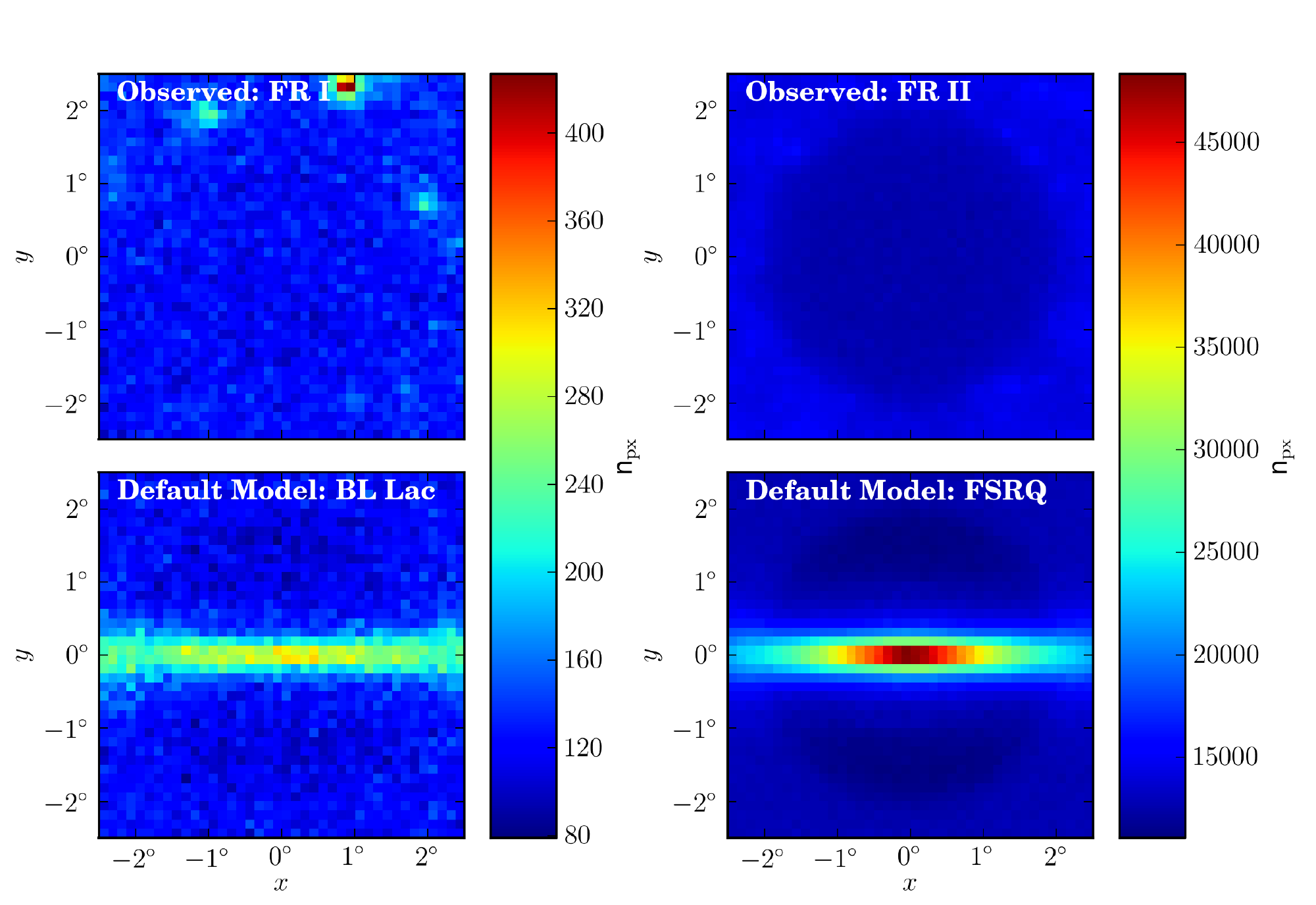}
  \end{center}
  \caption{Top: density of 1-100~GeV gamma rays in the stacked {\em Fermi} data after aligning the FR I (left) and FR II (right) samples based on their radio features. The combined front- and back-converted {\em Fermi} events are displayed.  To facilitate estimates of the Poisson noise per pixel, the total number of photons in each pixel is shown.  Bottom: the anticipated density of 1-100~GeV gamma rays arising from the IC halos from a stacked population of BL Lacs (left) and FSRQs (right), generated as described in the main text.  Note that these may be substantially reduced if an exponential cutoff in the intrinsic blazar spectra below 1~TeV is present or some other process preempts the formation of the IC halos.  In any case, the anticipated halo signal is clearly absent.  This remains the case when front- and back-converted events are considered separately.
  \label{fig:07}}
\end{figure*}

This process is repeated at least $5.5\times10^4$ times to produce an ensemble of results from which the expected statistical properties can be extracted.
Typically, this procedure produces $3\times10^4$ and $4\times10^6$ photons within $2^\circ$ for the FR I/BL Lac and FR II/FSRQ catalogs, composing 20\% and 30\% of the total photons within $2^\circ$ in the stacked images, respectively.  In both cases, this should be clearly visible in the stacked images.

\section{Stacked Gamma-ray Image Results} \label{sec:results}

Here we present the aligned and stacked {\em Fermi}-LAT images of the FR I and FR II sources selected in Section~\ref{sec:method}.  These are cocentered with and rotated in a fashion identical with that determined by the radio jet images.  The resulting images are presented for the stacked FR I and FR II samples in Section~\ref{sec:images}.  Various statistical characterizations of each in comparison to the anticipated IC halo signal are given in Sections~\ref{sec:images} and \ref{sec:ps}.  The large ensembles of simulated IC halos that we generate enable us to place significant limits on the existence of the IC halo signals.  In no instance is the observed gamma-ray distribution consistent with any realization, excluding IC halos associated with our canonical model, formally by more than 6$\sigma$, and likely much more.  We then investigate the impact of known systematic uncertainties in the IC halo simulations (Section~\ref{sec:systematics}).  Neither limiting the lifetime of radio jets, uncertanties in the gamma-ray SEDs, nor restricting gamma-ray-bright BL Lacs to low redshifts has a qualitative impact on this result.  The latter in particular is not relevant: FRICAT sources have $z<0.2$ and FSRQs are observed to be gamma-ray bright to high redshift \citep{FRICAT,3LAC}.

\subsection{Stacked Gamma-ray Images} \label{sec:images}
\begin{figure}
  \begin{center}
    \includegraphics[width=\columnwidth]{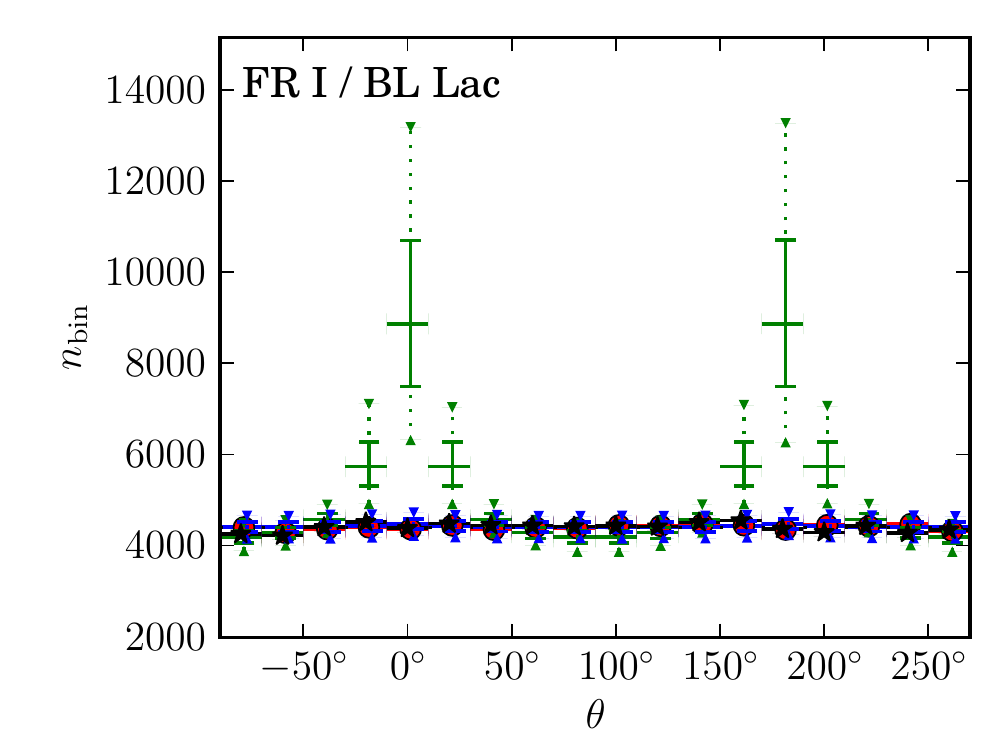}
    \includegraphics[width=\columnwidth]{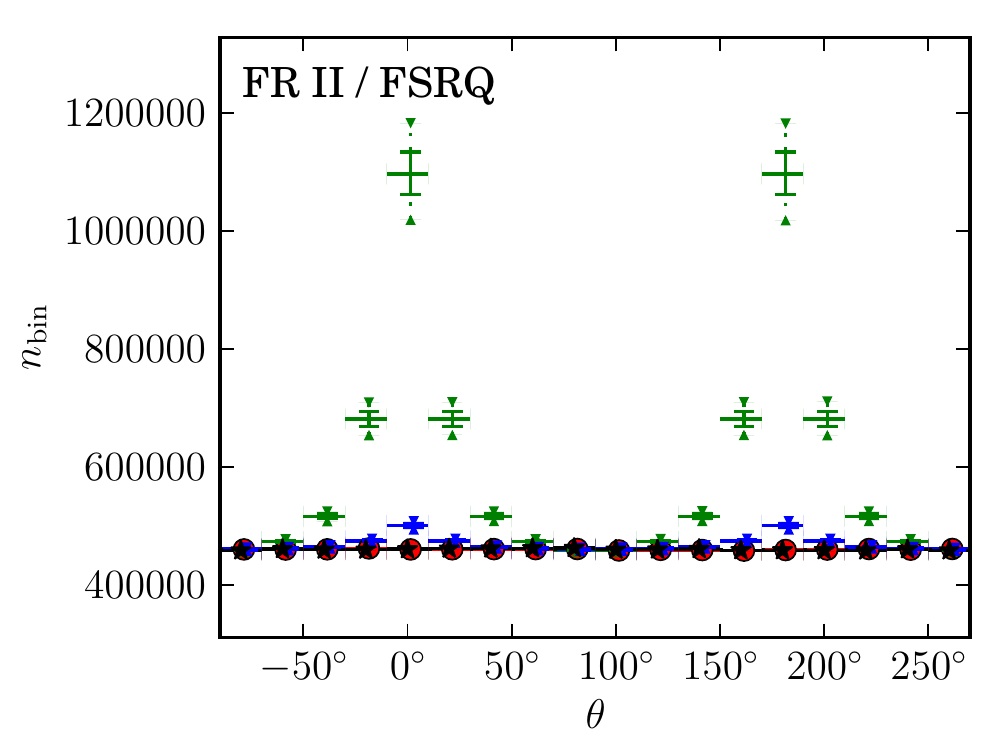}
  \end{center}
  \caption{Comparisons between the observed and expected angular distributions of gamma rays within $\circp{1}{8}$ in the stacked {\em Fermi} images of the FR I (top) and FR II (bottom) samples.  These are presented as angular histograms of the gamma-ray position angles about the stacked source locations.  In both plots, these are shown for the aligned (black stars) and randomly rotated (red circles) stacked cases.  The signals anticipated by the standard unified picture are shown by the green bars, for which the solid and dotted error bars indicate the one-sided 95\% and 99.99\% confidence regions, respectively.  The signals expected in the presence of beam-plasma instabilities are shown in blue.  Here we combine front- and back-converted events; considering each class separately produces similar results.  For clarity, points are horizontally shifted, and the dotted black line (not visible) shows the mean values of the number of photons per bin.
  \label{fig:08}}
\end{figure}

The resulting aligned and stacked gamma-ray images are shown in the top panels of Figure~\ref{fig:07}.  Absent in either stacked, aligned image is a clear, extended IC halo along the inferred jet direction.  Moreover, the stacked, aligned and control randomly rotated images are statistically indistinguishable.  Apparent in all of the stacked gamma-ray images is the presence of a weak, sudden rise in the flux just beyond $2^\circ$, corresponding to the presence of nonexcluded, bright gamma-ray point sources.  Through the PSF, this excess extends within $2^\circ$, resulting in the enhanced fluence at the boundary. In Figure~\ref{fig:08} we show the angular histograms, integrating the images in Figure~\ref{fig:07} radially, and in Figure \ref{fig:11} we show the angular power spectra of the stacked images for radii $r<\circp{1}{8}$ for each catalog type.

The nondetection of an extended, bi-lobed feature excludes our IC halo models at overwhelming significance.  For none of the more than $5.5\times10^4$ independent front- and back-converted stacked analyses do we anticipate a null result consistent with that observed.  As a result, we can exclude the default model by more than 6$\sigma$, limited only by the number of realizations explored.  This is a conservative estimate of the confidence level: the exclusion of the two jet sides has independent gamma-ray realizations for each, increasing the potential significance further, though we neglect this because of the intrinsic correlation induced by the identical underlying source realizations.  Bringing the median expected values into agreement with the mean of the observed angular distribution requires a reduction in the halo luminosity by a factor of 67 and 940 for the FR I/BL Lac and FR II/FSRQ comparisons, respectively.

\begin{figure}
  \begin{center}
    \includegraphics[width=\columnwidth]{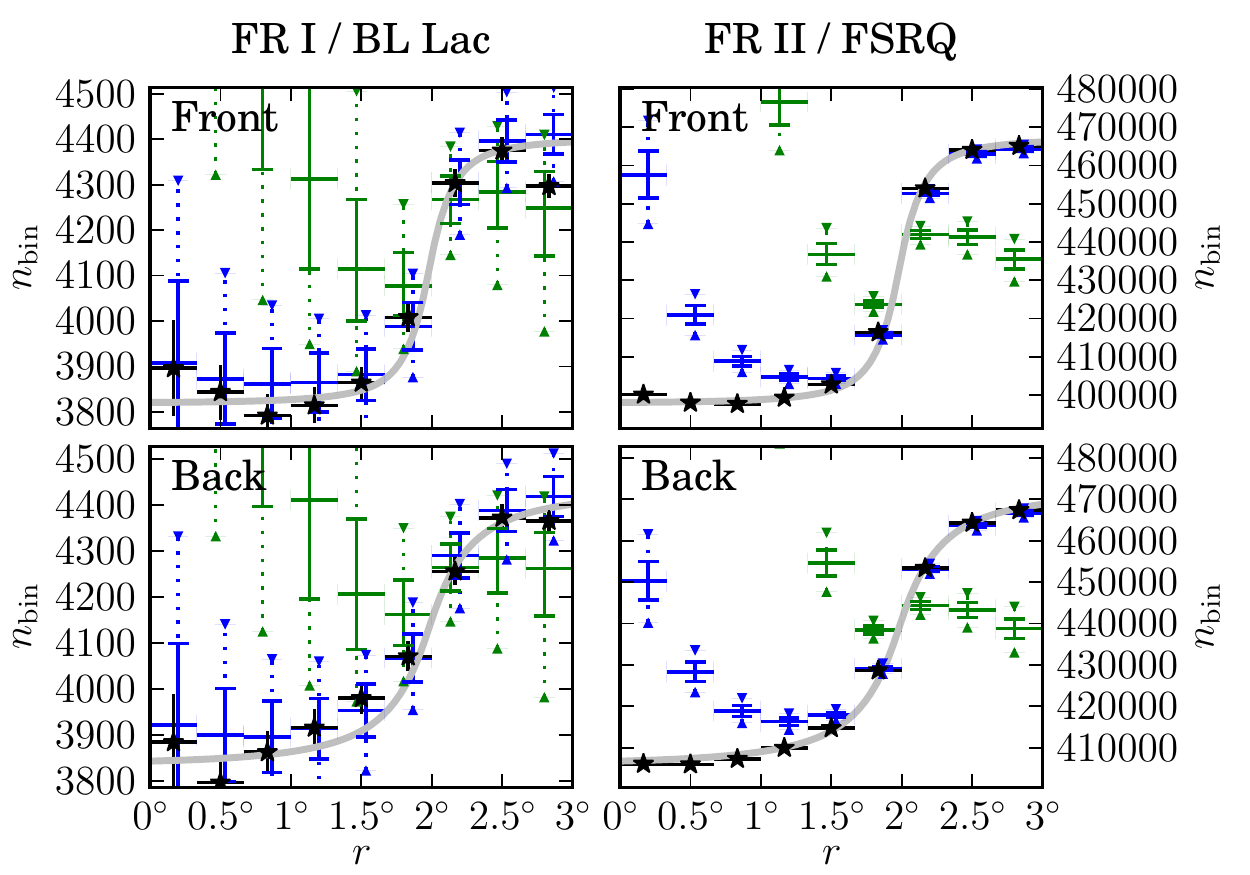}
  \end{center}
  \caption{Comparisons between the observed and expected radial distributions of gamma rays within $3^\circ$ in the stacked {\em Fermi} images of the FR I (left) and FR II (right) samples separately for front-converted (top) and back-converted (bottom) events.  These are presented as angular histograms of the gamma-ray position angles about the stacked source locations.  A step-function model, smoothed by the energy-averaged Pass 8R2\_V6 ULTRACLEANVETO PSF, is shown by the gray line.  The signals anticipated by the standard unified picture are shown by the green bars, for which the solid and dotted error bars indicate the one-sided 95\% and 99.99\% confidence regions, respectively.  The signals expected in the presence of beam-plasma instabilities are shown in blue.  For clarity, points are horizontally shifted, and the dotted black line (not visible) shows the mean values of the number of photons per bin.
  \label{fig:10}}
\end{figure}

The two-dimensional photon distributions shown in Figure~\ref{fig:07} exhibit a clear excess of gamma rays near the $2^\circ$ boundary.  This is a consequence of the exclusion of radio jet sources with a 3FGL source within $2^\circ$ coupled with the {\em Fermi} PSF.  That is, this excess is associated with 3FGL sources outside of the $2^\circ$ that bleed inside due to {\em Fermi}'s finite angular resolution.  To demonstrate this explicitly, in Figure~\ref{fig:10} we show the radial distribution of the photon density (note that this is independent of source alignment).  This is compared to a simple model, consisting of separate uniform densities inside and outside the $2^\circ$ convolved with an appropriate PSF.  The values for these densities are set by the interior and exterior photon densities.  Because the {\em Fermi} Pass 8R2\_V6 ULTRACLEANVETO PSF is energy dependent, we average over an SED consistent with that observed in the background (and consistent with both the interior and exterior photon populations), $dN/dE\propto E^{-2.38}$.  These are shown in Figure~\ref{fig:10} for front- and back-converted events separately because of the rather different {\em Fermi} PSFs for each set of photons.

The step-function model does a good job of reproducing the excess just inside $2^\circ$ and the corresponding dearth just outside $2^\circ$.  There are weak indications of a slight second excess of photons in the front-converted events at the field centers, associated with the double radio sources themselves.  It is tempting to interpret this as a tantalizing signal of gamma-ray emission associated with radio lobes.  However, there is no evidence for a point source in the back-converted events (lower panels of Figure~\ref{fig:10}); importantly, even with the larger PSF associated with back-converted events, a point source with the flux implied by the front-converted excess would be clearly visible.  In comparison, our default simulation would exhibit a substantial, centrally concentrated excess associated with the IC halo emission that is clearly not present.  Note that this excludes the possibility that we have inadvertently simply misaligned the stacked sources.

\subsection{Angular Power Spectra} \label{sec:ps}
\begin{figure}
  \begin{center}
    \includegraphics[width=\columnwidth]{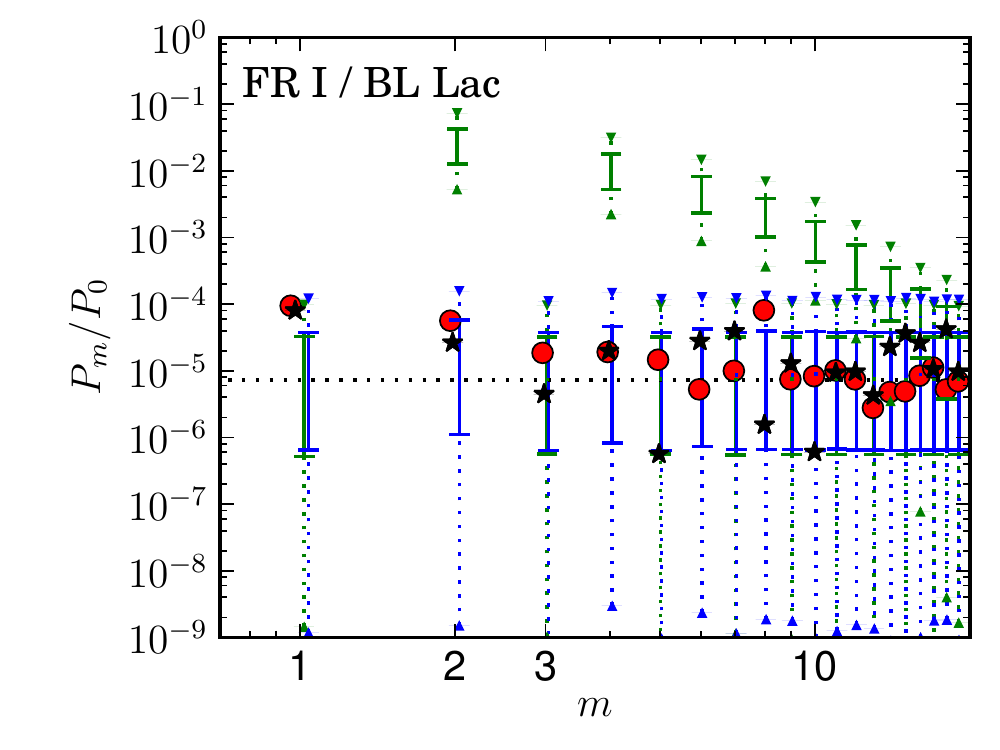}
    \includegraphics[width=\columnwidth]{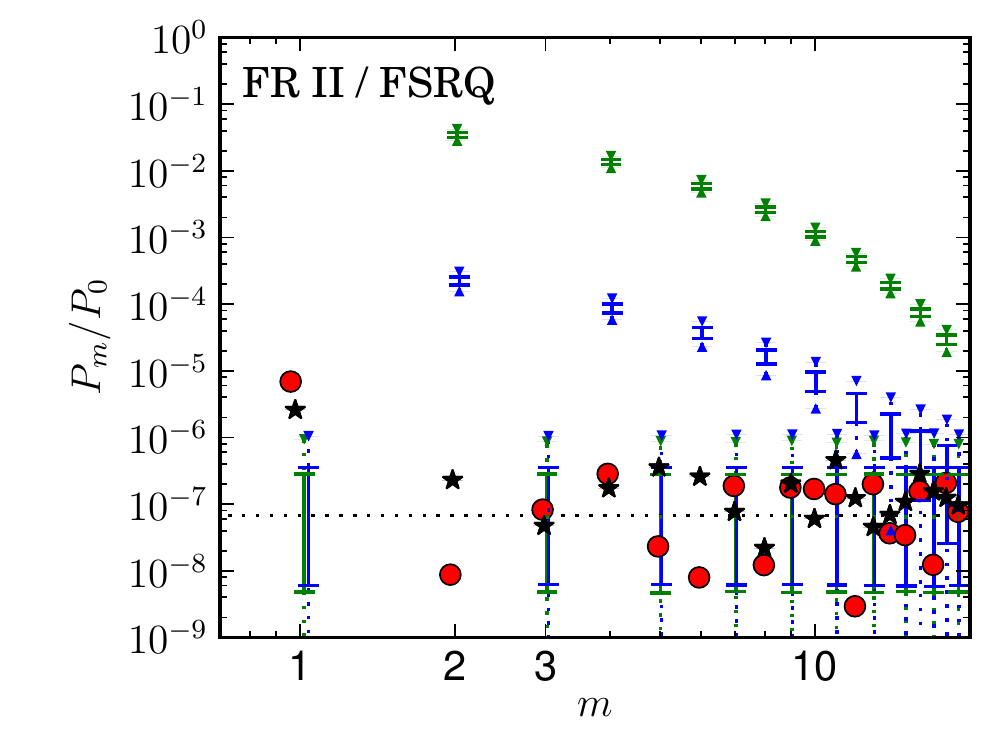}
  \end{center}
  \caption{Comparisons between the observed and expected angular power spectra of gamma rays within $\circp{1}{8}$ in the stacked {\em Fermi} images of the FR I (top) and FR II (bottom) samples.  In both plots, these are shown for the aligned (black stars) and randomly rotated (red circles) stacked cases.  The signals anticipated by the standard unified picture are shown by the green bars, for which the solid and dotted error bars indicate the one-sided 95\% and 99.99\% confidence regions, respectively.  The signal expected in the presence of beam-plasma instabilities is shown in blue.  Here we combine front- and back-converted events; considering each class separately produces similar results.  For clarity, points are horizontally shifted, and the dotted black line shows the Poisson limit.
  \label{fig:11}}
\end{figure}

Angular power spectra provide an alternative means of assessing the anisotropy of the stacked images that has been extensively discussed previously  \citep{BTII,BTIII}.  While angular power spectra present a powerful tool for stacking unaligned images, i.e., incoherent stacks, here we employ them to characterize the stacked, aligned images, i.e., coherent stacks, to provide a natural connection to prior work. These are shown in Figure~\ref{fig:11} for the two samples. For comparison, the anticipated angular power spectra are also shown. Generally, similar trends are apparent as seen in the angular histograms: the default halo model is excluded at overwhelming significance. Unlike the angular histogram, the angular power spectra indicate the degree to which {\em any} bi-lobed feature is excluded, independent of the particular alignment precision.

\subsection{Potential Systematic Uncertainties} \label{sec:systematics}
Here we explore a variety of potential systematic uncertainties and explore their potential impact on the nondetection of IC halos. First, we explore instrumental uncertainties from the {\em Fermi}-LAT. That is, using the front and back detectors, we compare our results to ensure that the results in both are consistent, i.e., that neither is biasing our result. Second, we analyze the impact of the finite lifetime of radio/gamma-ray jets. Finally, we analyze the impact that a potential spectral curvature at high energies ($\sim 100\rm GeV-1\rm TeV$) from a variety of sources may have.  We find that none of these are capable of fully explaining the absence of IC halos.

\subsubsection{Front- vs. Back-converted Event Comparison}
\begin{figure*}
  \begin{center}
    \begin{tabular}{cc}
      \includegraphics[height=0.5\textwidth]{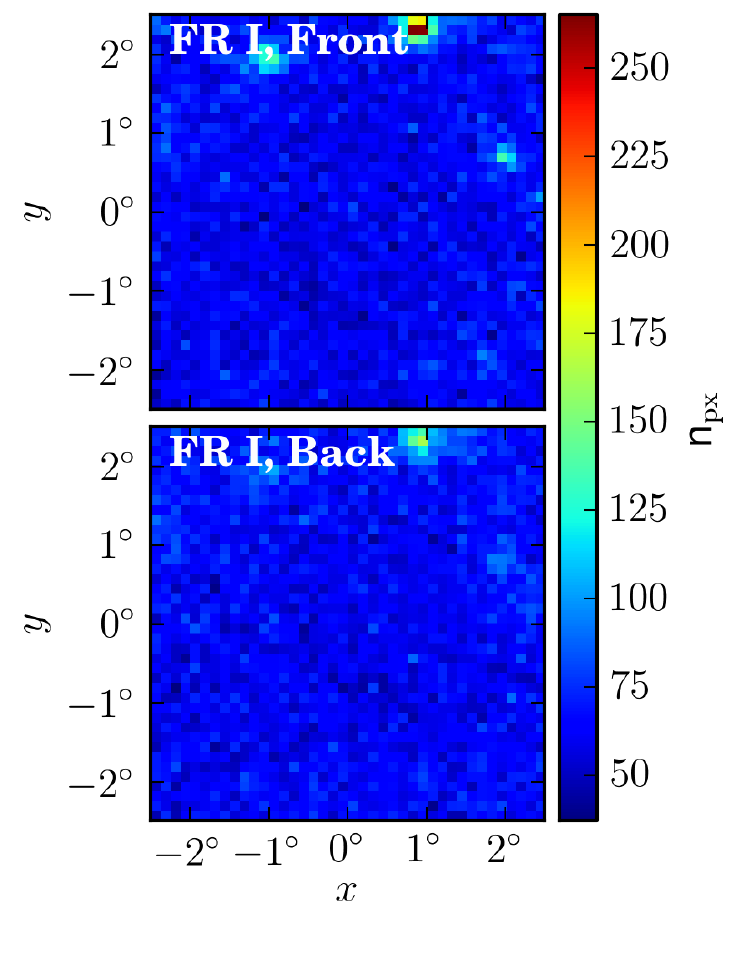}
      & \includegraphics[height=0.5\textwidth]{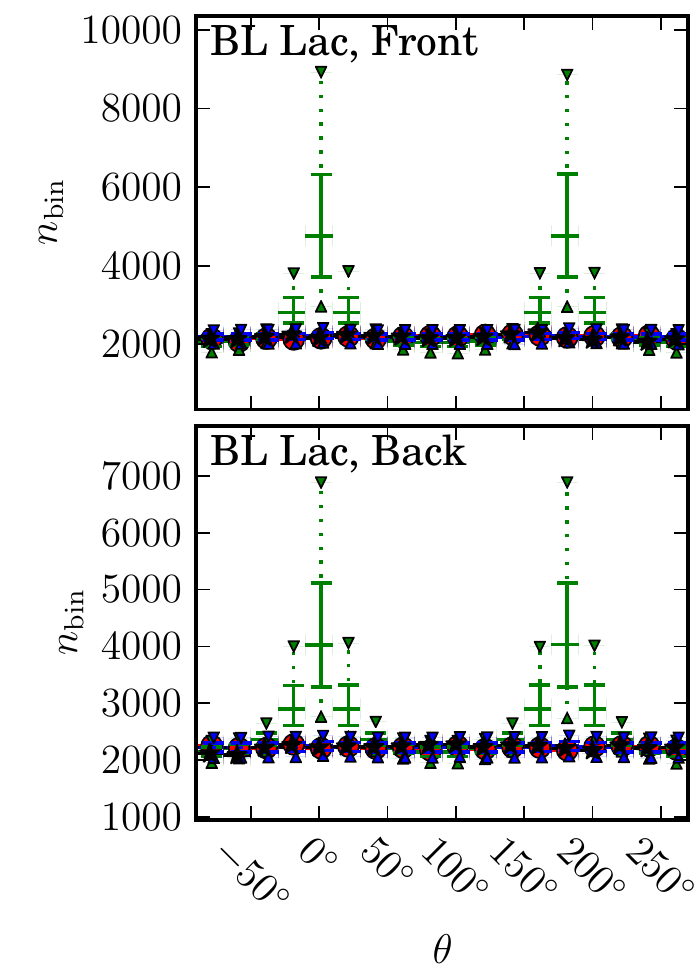}\\
      \includegraphics[height=0.5\textwidth]{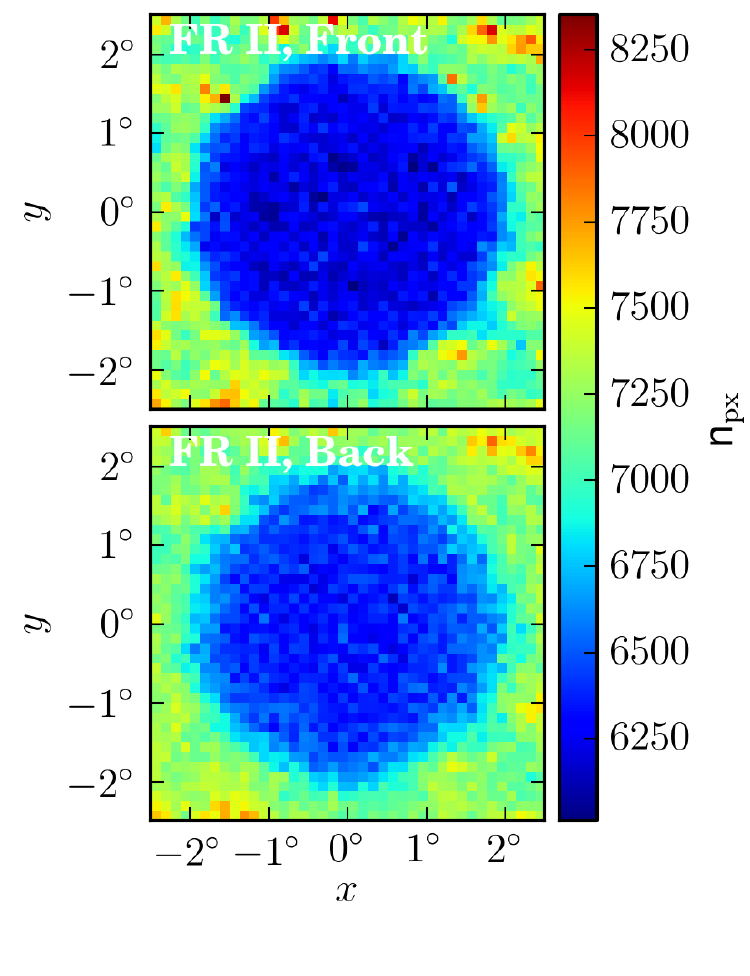}
      & \includegraphics[height=0.5\textwidth]{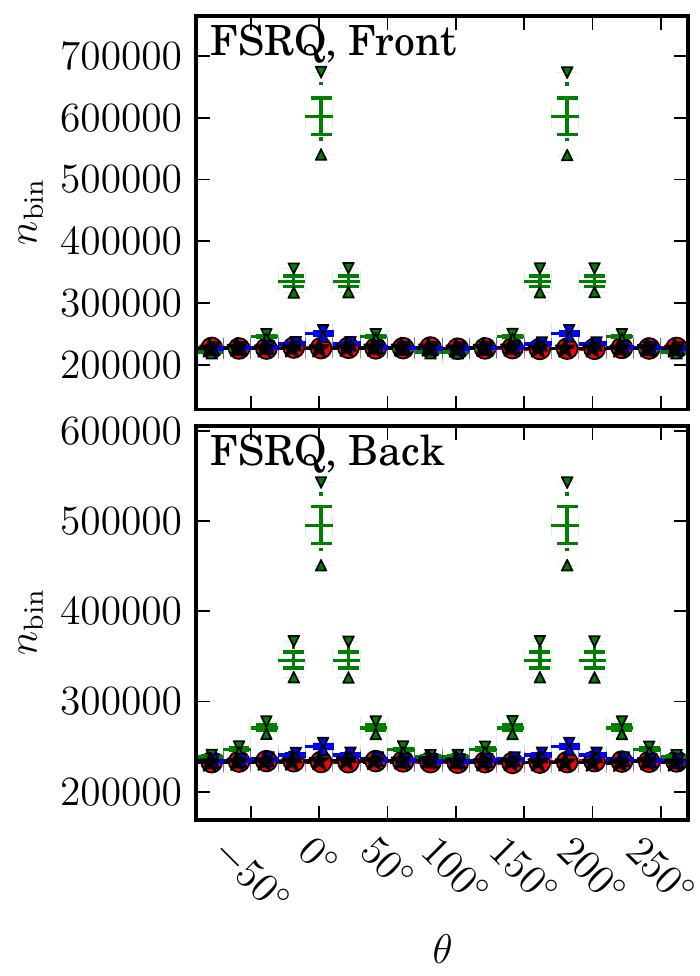}
    \end{tabular}
  \end{center}
  \caption{Comparisons of front- and back-converted stacked gamma-ray maps for the FR I/BL Lac and FR II/FSRQ analyses, organized by row.  Left: density of 1-100~GeV photons after alignment for the indicated stacked sample and Pass 8R2\_V6 ULTRACLEANVETO event class.  Right: associated angular distributions of the photons within $\circp{1}{8}$ after alignment (black stars) and random rotations (red circles), compared with the anticipated signals from the standard unified picture (green bars) and when beam-plasma instabilities are included (blue bars).  Solid and dotted error bars indicate the one-sided 95\% and 99.99\% confidence regions, respectively.  For clarity, points are horizontally shifted, and the dotted black line (not visible) shows the mean values of the number of photons per bin.
    \label{fig:12}}
\end{figure*}

In practice, we construct each of the comparisons independently for front-converted and back-converted events, combining them only at the end to improve the statistical significance of the result.  In no case do we see statistically significant disparities between the two sets of data.  For completeness, we show these comparisons here.  Figure~\ref{fig:12} shows the two-dimensional histograms of the two different data sets independently and the corresponding angular histogram comparisons.  As with the combined presentations, these are independently excluded at more than 4$\sigma$.  As a result, the joint exclusion is more than 6$\sigma$.

\subsubsection{Impact of Radio Lifetimes}
\begin{figure*}
  \begin{center}
    \begin{tabular}{cc}
      \includegraphics[width=0.48\textwidth]{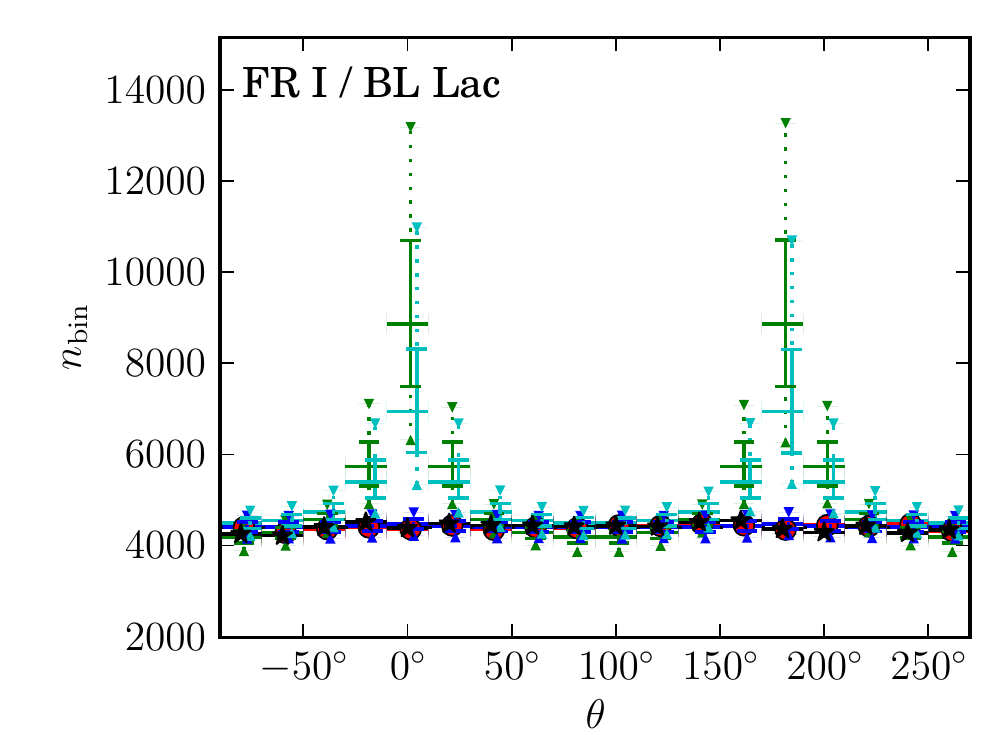}
      &
      \includegraphics[width=0.48\textwidth]{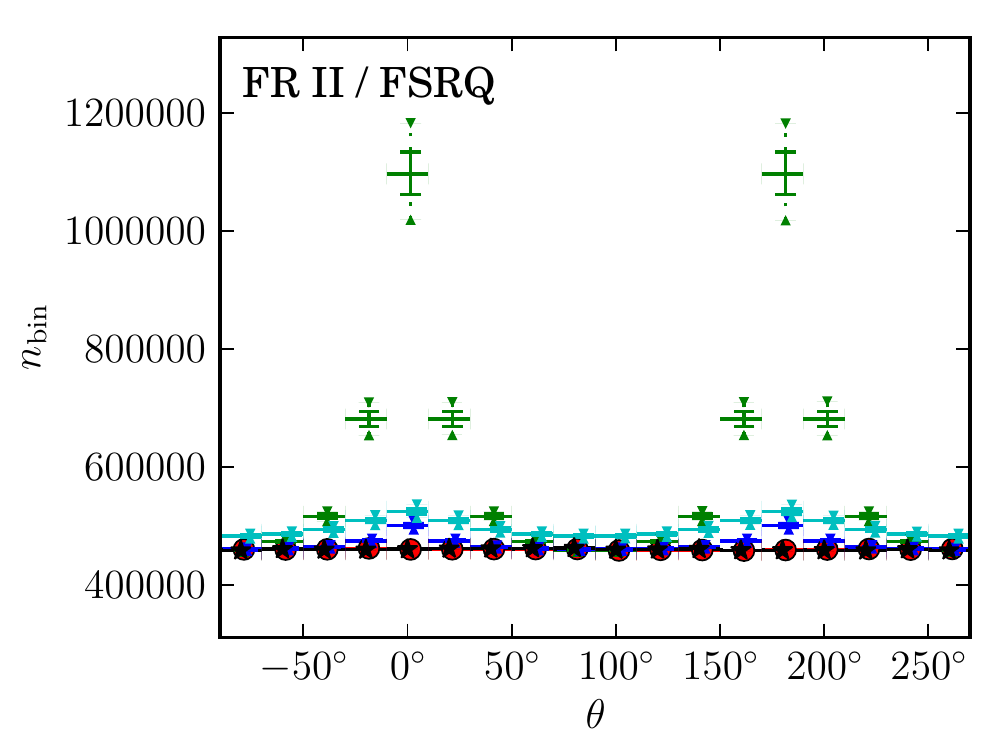}\\
      \includegraphics[width=0.48\textwidth]{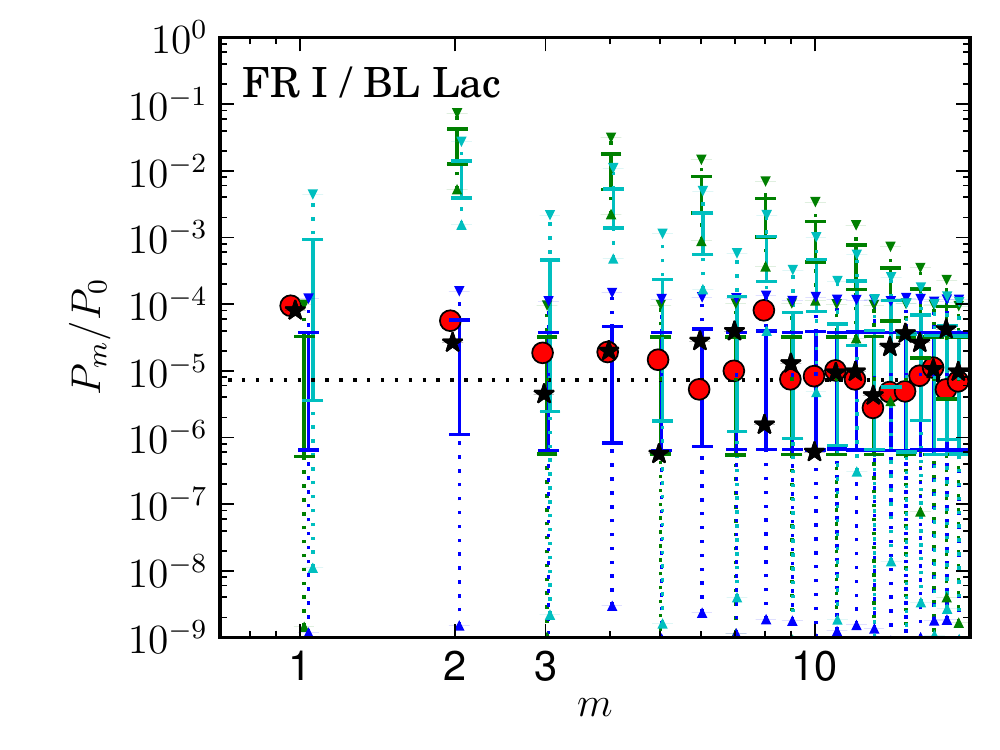}
      &
      \includegraphics[width=0.48\textwidth]{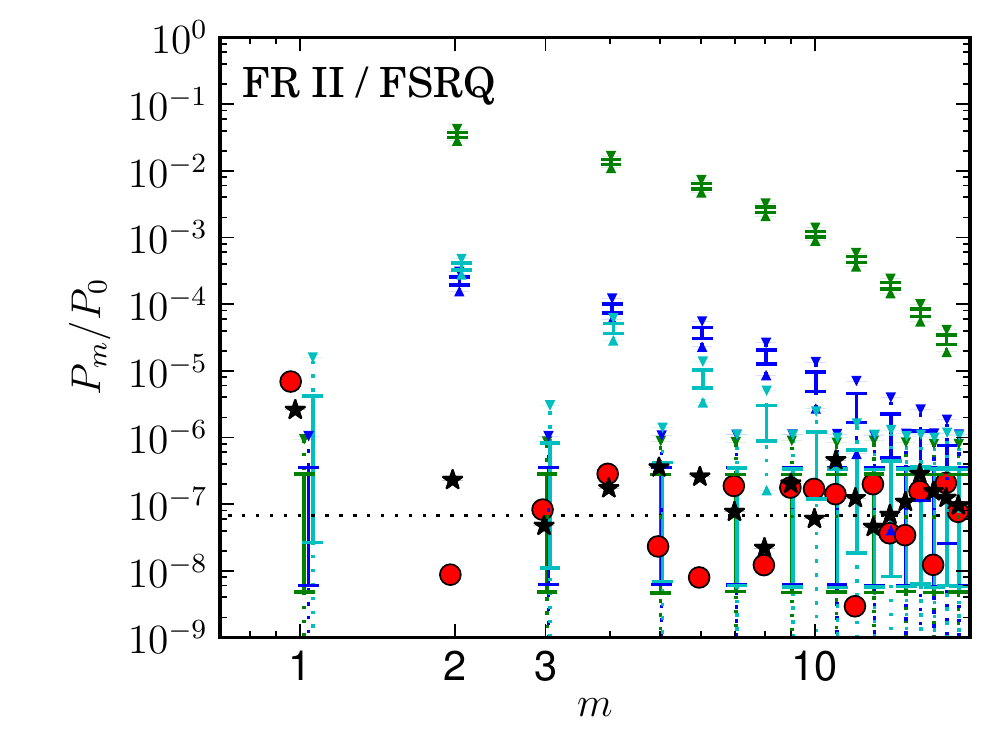}
    \end{tabular}
  \end{center}
  \caption{Comparisons between the observed and expected angular distributions of gamma rays within $\circp{1}{8}$ in the stacked {\em Fermi} images of the FR I (left) and FR II (right) samples after limiting the jet lifetimes to 30~Myr.  These are presented as angular histograms (top) and angular power spectra (bottom) of the gamma-ray position angles about the stacked source locations.  In both plots, these are shown for the aligned (black stars) and randomly rotated (red circles) stacked cases.  The signals anticipated by halos associated with jets with lifetimes limited to 30~Myr is shown by the cyan bars, for which the solid and dotted error bars indicate the one-sided 95\% and 99.99\% confidence regions, respectively.  For reference, the signals expected by our default halo simulation and in the presence of beam-plasma instabilities are shown in green and blue, respectively.  Here we combine front- and back-converted events; considering each class separately produces similar results.  For clarity, points are horizontally shifted, and the dotted black line shows the mean values of the number of photons per bin.
  \label{fig:13}}
\end{figure*}

In our default and plasma cooled models, we have assumed that the VHEGR jets persist with the same orientation over the entire time the visible halo is generated.  While VHEGR sources are known to flare on short timescales, implying that they can turn on and off rapidly, it is the VHEGR flux averaged over the typical IC cooling time, roughly $2.4(1+z)^{-4}$~Myr, that we require to be long lived.  It is natural to assume that the VHEGR and radio jets are contemporaneous features of AGNs.  However, a conservative limit on the timescale of the cooling-time-averaged VHEGR emission epoch would then arise from the cooling timescales of the radio features.  To assess the impact of restricting the IC halo formation to this timescale, we have generated a set of simulated halos with a limited central-engine lifetime.  That is, we define a timescale as a function of halo angular extent, including the time of flight of the VHEGRs and subsequent IC halo photons, $t=[D_A(z)/(c\sin\Theta)]\theta(1\pm\cos\Theta)$, which we then restrict to be less than 30~Myr, the typical lower limit on the jet lifetime from the synchrotron cooling of the radio emission regions.  When confined to the $2^\circ$ window of interest, this excludes 40\% and 78\% of photons in our default FR I/BL Lac  and FR II/FSRQ models, respectively.  As seen in Figure~\ref{fig:13}, alone this is insufficient to materially alter our conclusion that IC halos remain excluded at overwhelming significance.

\subsubsection{Curvature of VHEGR SED} \label{sec:curve}
The IC halo flux depends linearly on the underlying TeV luminosity distributions of the gamma-ray jets.  While we have made an effort to conservatively estimate these and propagate the uncertainties of our estimates through our IC halo simulations, our redshift-luminosity distribution is predicated on the applicability of the convex broken-power-law SED model to the intrinsic gamma-ray SEDs of blazars between 100~MeV to 2~TeV.  While the reported fluxes in the 3FHL extend to 2~TeV, only a subset of sources are detected at such high energies, due either to the rapid decrease in photon flux associated with power-law SEDs even for hard sources or the absorption on the infrared background for objects with $z>0.2$.  For these reasons, only 33\% of the BL Lacs with redshifts and 8\% of the FSRQs with redshifts are detected above 150~GeV.  Thus, it remains possible, in principle, that an extreme curvature of the intrinsic VHEGR SED above 100~GeV may reduce the TeV luminosities significantly from our fit estimates.

It is clear, however, that a modest break, such as a cooling break, in the SED is insufficient.  Such a break has been inferred from comparisons between the photon spectral index at 1~TeV and 1~GeV of {\em Fermi} blazars \citep{3LAC}, typically resulting in $\Gamma_{\rm TeV}-\Gamma_{\rm GeV}\approx 1$ \citep{2LAC}; the break between the 100~GeV and 1~TeV photon spectral index is smaller generally.  Above 100~GeV, this would provide an insufficient energy range in which to reduce the expected halo signal sufficiently to be consistent with a nondetection.  That is, a far more aggressive reduction in the intrinsic emission is required, e.g., an exponential cutoff.

There is no empirical evidence for such cutoffs in the VHEGR SEDs of nearby BL Lacs and FSRQs after deabsorption.  For example, for all seven of the FSRQs listed in TeVCat (S3~0218+35, PKS~0736+017, TON~0599, 4C~+21.35, 3C~279, PKS~1441+25, PKS~1510-089; \citealt{TeVCat:2008}\footnote{See {\tt http://tevcat.uchicago.edu}, Catalog Version 3.4}), no high-energy cutoffs are seen during flaring episodes; any emission model that exhibits a rapid cutoff must avoid doing so during active periods.  This is shown explicitly in Figure~\ref{fig:04} for two of the TeVCat FSRQs.  For BL Lacs, the case is even clearer, due to the higher number of nearby, hard objects, for which the SED is measured above 100~GeV directly.

Similarly, there are theoretical reasons to not expect a spectral cutoff between 100~GeV and 1~TeV for both leptonic and hadronic emission models.  Cutoffs at high energies can arise in leptonic models from two sources: the Klein-Nishina suppression of the Compton cross section \citep{BlumenthalGould1970} or an intrinsic superexponential cutoff in the underlying lepton population \citep{Zirakashvili07}.  The Klein-Nishina suppression occurs when in the lepton frame the seed photon energies are comparable to the electron rest mass, that is, when $E_{\rm seed} E_{\gamma} \approx 2 m_e^2 c^4$, where $E_\gamma$ and $E_{\rm seed}$ are the energies of the IC gamma ray and the original seed photon, respectively.  This results in a break in the gamma-ray SED above an energy $E_\gamma\approx 500 (E_{\rm seed}/1~{\rm eV})^{-1}~\mathrm{GeV}$.  However, even for very extreme seed photon distributions, this is modest and generally subexponential.  For example, for a Plankian seed photon spectrum that peaks at 5~eV being IC scattered by a power-law lepton distribution $\propto E^{-2.1}$, the resulting gamma-ray SED is $\propto E^b \exp[-(E/E_c)^{0.25}]$ for $E_c=2$~GeV, which is considerably lower than the estimated value of $E_\gamma$.  Above $E_c$, this is quite flat in the energy spectrum $E L_E$, reminiscent of high-synchrotron-peak (HSP) blazars, and being dominated by a subexponential cutoff near a TeV.  Seed photon populations that peak at lower energies have corresponding higher cutoffs, and thus adopting a more broadly distributed seed photon distribution, such as the double power laws associated with the synchrotron peaks in blazars, reduces the impact of the Klein-Nishina suppression further.

Alternatively, the IC-scattering lepton populations themselves may exhibit an intrinsic cutoff.  For diffusive shock acceleration (DSA) in nonrelativistic shocks\footnote{Blazar shocks are relativistic, but the maximum energy is determined by similar processes, so it is instructive to look at the well-studied X-ray synchrotron and IC gamma-ray spectra from nonrelativistic astrophysical shocks.}, applicable to supernovae remnants, the shape of the lepton distribution changes when the energy of the accelerated leptons reaches a maximum determined by the competition between DSA and radiative losses.  In the cutoff region, the lepton momentum distribution function is proportional to a modified power-law term and a superexponential term, $\exp(-p^2/p_0^2)$, where $p$ is the magnitude of the momentum and $p_0$ is a scale set by the competing timescales.  The former reflects a pileup of leptons as their cooling time becomes comparable to the acceleration time.  The exponential term, however, effectively cancels this pileup feature, which results in a prolonged power law up to the electron cutoff momentum, where a steeper superexponential cutoff takes over \citep[see Fig.~3 of][]{Zirakashvili07}.  Because the lepton momentum for relativistic particles is $p\approx\gamma m_e c$, a superexponential cutoff in the lepton distribution produces an exponential cutoff in the IC-scattered gamma rays of $\exp(-E/E_c)$ with a scale of $E_c=2 E_{\rm seed} p_0^2/m_e^2 c^2$.  The scale of the lepton momentum cutoff, $p_0$, and thus $E_c$, depends most strongly on the strength of the shock and the ambient magnetic field strength.  However, for all credible values of the shock velocities and magnetic field strengths in blazar jets and their environments, this results in $p_0/m_e c\gg10^8$ and thus $E_c\gg1$~TeV \citep[see Eq.~22 of][]{Zirakashvili07}.

Hadronic models for the GeV-TeV emission typically arise from an underlying proton population that extends to EeV energies, limited ultimately by the smaller of the two following values: the reachable maximum energy according to the modified Hillas criterion, $E_{\mathrm{max}}=ZeBRv_{\mathrm{shock}}/c$, where $Ze$ is the particle charge, $B$ is the shock magnetic field strength, $R$ is the shock confinement radius, and $v_{\mathrm{shock}}$ is the shock velocity, or the GZK cutoff \citep{Greisen1966,ZatsepinKuzmin1966}. The recent detection of a 290~TeV neutrino associated with blazar TXS 0506+056 \citep{TXSblazarI,TXSblazarII} provides evidence for photo-pion emission from energetic protons and, therefore, the existence of similarly high energy photons.  This has now been interpreted in terms of hybrid leptohadronic models for TXS 056+056, in which the sub-TeV emission is dominated by leptonic processes, transitioning above a TeV to hadronic origins, with an associated concave spectral break in the intrinsic spectrum \citep[see, e.g.,][though note that the SEDs shown there are after absorption]{Ahnen2018,Cerruti2018,Gao2018}.  Thus, relative to these hybrid models, our convex broken power-law fits will {\em systematically underestimate} the TeV luminosities of similar objects.

As a result, we conclude that it is unlikely that our TeV luminosity estimates are larger by more than an order of magnitude than the true values, and thus an additional plausible spectral curvature alone is unable to explain the nondetection of IC halos in either the FR I/BL Lac or FR II/FSRQ comparisons in the absence of novel radiative processes. In combination with limiting VHEGR jet lifetimes to 30~Myr, a modest spectral curvature can explain the apparent absence of an IC halo signal in the stacked FR II objects but cannot do so for the stacked FR I objects.

\section{Discussion} \label{sec:discussion}

\begin{figure}
  \begin{center}
    \includegraphics[width=\columnwidth]{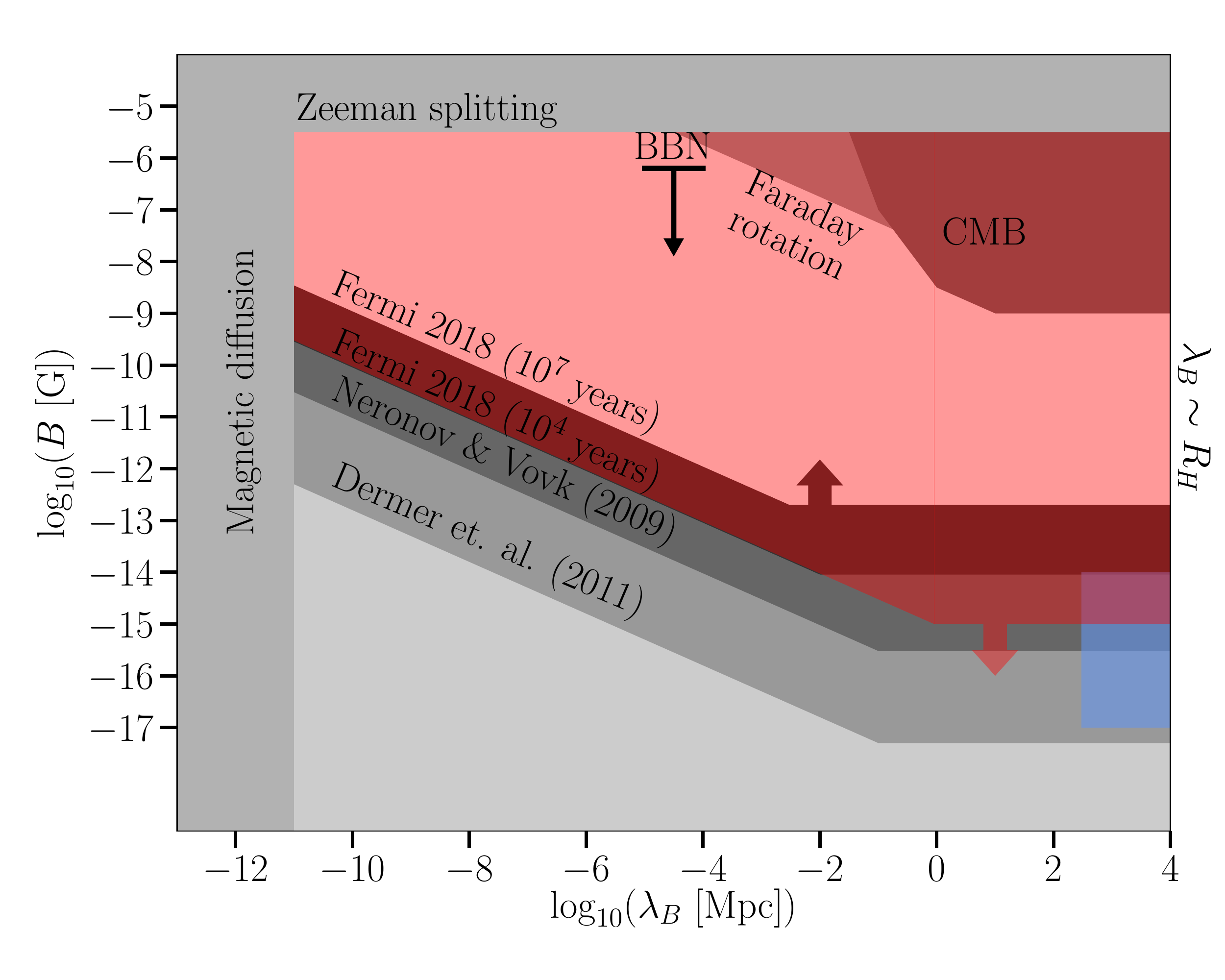}
  \end{center}
  \caption{Summary of IGMF constraints on strength and correlation length arising from the nondetection of IC halos described here in comparison to various others. The red shaded area shows the exclusion region from nondetection of stacked anisotropic gamma-ray halos (this work).  The outer gray area indicates regions excluded by magnetic diffusion (left) and direct limits from Zeeman splitting (top).  Beyond the right axis, $\lambda_B$ is larger than the Hubble radius, $R_H$.  The inferred lower limits from 1ES 0229+200 \citep{Nero-Vovk:10} and the modified lower limit after allowing for extreme limits on TeV emission lifetime \citep{Derm_etal:11} are shown by the lower two shaded regions.  The lower limits arising from morphological limits for TeV-bright sources assuming gamma-ray jet lifetimes of $10^4$~yr and 10~Myr are shown in dark gray \citep{FermiIGMF:18}.  The blue region shows the region excluded by searches for IC halos about bright {\em Fermi} sources \citep{BTIII}. Note that the colored and gray regions are exclusion regions and there is no allowed region (white) left.  Combined, these constraints rule out a volume-filling IGMF for all of the IGMF parameter space, including $\lambda_B>R_H$ (and leaves the absence of IC features in {\em Fermi} blazars unexplained).  Alternatively, it argues for the existence of an additional physical process that preempts the IC cascade in the form of, for example, virulent beam-plasma instabilities that dominate the energy loss of TeV pairs. (Adapted from \citealt{Nero-Vovk:10}.)
    \label{fig:09}}
\end{figure}

An IGMF with strength less than $1\times10^{-15} \eta^{-1} (\lambda_B)$~G would avoid the extended, anisotropic gamma-ray halos excluded here.  Thus, the nondetection presented here may be interpreted narrowly as an upper limit on the strength of a putative IGMF, shown in Figure~\ref{fig:09}.
When combined with limits from the SEDs of TeV-blazars \citep{Nero-Semi:09,Nero-Vovk:10,Tayl-Vovk-Nero:11,Taka_etal:11,Vovk+12}, this would leave a narrow region of permitted IGMF strengths that depends on $\lambda_B$.  Above $\lambda_B\approx0.9~{\rm Mpc}$, this is limited to between $3\times10^{-16}$~G and $10^{-15}$~G.  For smaller $\lambda_B$, which are consistent with well-motivated models of magnetogenesis, the required IGMF strength grows as $\eta^{-1}(\lambda_B)$ \citep[for fields generated at the electroweak or quantum chromodynamics phase transitions, the correlation length is below 10~kpc, though it may grow to 0.1~Mpc over the age of the universe; see][]{2008RPPh...71d6901K,2012SSRv..166....1R,2012SSRv..166...37W}.
However, this is already in direct conflict with morphological searches for the GeV halo IC component of known TeV sources: when jet lifetimes in excess of $10^4$~yr are considered, IGMF strengths greater than $10^{-14}$~G for $\lambda_B\approx0.01$~Mpc, scaling as $\lambda_B^{-1/2}$ for smaller correlation lengths, are required \citep{FermiIGMF:18}.  Therefore, {\em there is no consistent explanation for the simultaneous apparent absence of line-of-sight IC halo emission and that associated with oblique radio jets}, and we do not consider this possibility further, focusing on the implications for the formation of the IC halos themselves.

Interpretations of the absence of the IC halos may then be organized into two general categories: either the VHEGR emission responsible for generating the progenitor electron-positron pairs is missing, or some other process dominates their cooling subsequent to their formation.  Both of these require novel processes that fundamentally modify our understanding of the extragalactic gamma-ray universe.  The first is equivalent to assuming that our inferred TeV luminosities may be biased, either because radio-bright jets are not the proper parent population of objects or, as we argue in Section~\ref{sec:curve}, some as-yet unidentified process suppresses the TeV emission of gamma-ray-bright blazars.  In the absence of such an extreme spectral modification for source classes of FSRQs and BL Lacs, we turn to the remaining explanation.

This requires a mechanism by which streaming populations of relativistic electron-positron pairs will lose energy on timescales considerably shorter than $2.4 (1+z)^{-4}$~Myr within cosmological voids.  A natural explanation for this is the presence of virulent beam-plasma instabilities in the IGM, associated with the anisotropic nature of the pairs generated by the original VHEGR emission \citep{PaperI}.  These suppress the IC halo by an amount that depends on the relative cooling rates associated with saturated plasma instability and IC cooling.  In the linear regime, the instability cooling rates are large in comparison to the IC cooling timescale for all relevant redshifts \citep{PaperI}.  This is modestly reduced by the presence of nonlinear Landau damping \citep{Chang:2014}, though the nonlinear saturation remains poorly understood \citep{Shalaby:resol,Shalaby:inhomogeneity,Vafin18}, and thus the linear instability cooling rates are a credible upper limit on the impact of the plasma instabilities on the formation of IC halos.  For the FR I/BL Lac population, the plasma cooling is capable of fully explaining the lack of large-scale IC halos.  As seen in the bottom panel of Figure~\ref{fig:11} and the right panels of Figure~\ref{fig:13}, for the FR II/FSRQ, an additional reduction, in the form of either an enhanced cooling or suppression of the FSRQ VHEGR emission is required; a reduction by an order of magnitude in the TeV luminosities of FSRQs is sufficient and requires only a modest evolution in the SED above 100~GeV that is broadly consistent with the gamma-ray SEDs of {\em Fermi} sources \citep{3LAC}.

\section{Conclusions} \label{sec:conclusion}
We have identified, aligned, and stacked the gamma-ray images of the oblique radio analogs of the gamma-ray-bright blazars.  This was done independently for FR I and II objects, which we compared to the expectations from simulated IC halos from stacked BL Lacs and FSRQs, respectively.  Based on this, we can conservatively exclude the existence of IC halos at more than 6$\sigma$.  The apparent absence of IC halos cannot be explained via identified systemic uncertainties in the analysis.

Alone this requires an IGMF less than $1\times10^{-15} \eta^{-1} (\lambda_B)$~G. Combined with prior constraints from {\em Fermi} blazar observations, which limit the IGMF to be greater than $3\times10^{-13}$~G (and progressively larger in the diffusive regime when the correlation length of the IGMF becomes smaller than the IC cooling time) for VHEGR jet lifetimes consistent with those assumed here, this precludes any interpretation of this nondetection within the context of an IGMF.  That is, there is no IGMF that would simultaneously explain all nondetections.  This substantially complicates efforts to probe the IGMF with gamma-ray observations.

This suggests either (1) a novel process that suppresses the VHEGR emission dramatically between 100~GeV and 1~TeV, for which there is currently little empirical or theoretical support, or (2) a mechanism by which the IC halos are preempted after the relativistic pairs are generated by VHEGR absorption on the infrared background.  Either of these solutions requires fundamental revisions to our understanding of the origin and/or impact of the VHEGR emission of AGNs.  Modifications to the emission necessarily imply additional, possibly redshift-dependent, processes within the gamma-ray emission regions of AGNs, placing a severe constraint on their nature and location.  Additional cooling processes in the IGM reprocess the VHEGR luminosity of AGNs into forms other than GeV halos, which are essentially decoupled from all other components in the IGM.  Beam-plasma instabilities may serve this role \citep{PaperI,Chang:2014}, draining energy from the relativistic pairs into plasma waves and, ultimately, heat.  If these instabilities are efficient, they would serve as a mechanism that satisfies (2) and, thus, invalidates the current IGMF limits from the nonobservation of GeV halos around TeV blazars.  Additionally, beam-plasma instabilities deposit this energy into heat in the IGM, dominating the energy budget in cosmic voids, where it can raise the IGM temperature by up to an order of magnitude at $z=0$ \citep{PaperII,PaperIV,Lamberts:2015}, substantially modifying the Lyman-$\alpha$ forest at late times ($z\lesssim 2$), and possibly suppressing late-time star formation in galaxies, especially dwarfs \citep{PaperIII}.

\acknowledgments
  A.E.B., P.T.,~and M.S.~receive financial support from the Perimeter
Institute for Theoretical Physics and the Natural Sciences and
Engineering Research Council of Canada through a Discovery Grant and
through the Alexander Graham Bell scholarship (P.T.).  Research at the
Perimeter Institute is supported by the Government of Canada through
Industry Canada and by the Province of Ontario through the Ministry of
Research and Innovation.  A.E.B. thanks the Delaney family for their generous financial support via the Delaney Family John A. Wheeler Chair at Perimeter Institute.
C.P., M.S., and M.W.~acknowledge support by the European
Research Council under ERC-CoG grant CRAGSMAN-646955. P.C.~gratefully
acknowledges support from NSF grant AST-1255469. A.L.~receives
financial support from an Alfred P. Sloan Research Fellowship, NASA
ATP grant NNX14AH35G, and NSF Collaborative Research Grant 411920 and
CAREER grant 1455342. E.P.~acknowledges support by the Kavli
Foundation.

\bibliographystyle{aasjournal_aeb}
\bibliography{bt4bib}

\begin{thebibliography}{}
\expandafter\ifx\csname natexlab\endcsname\relax\def\natexlab#1{#1}\fi
\providecommand{\url}[1]{\href{#1}{#1}}
\providecommand{\dodoi}[1]{doi:~\href{http://doi.org/#1}{\nolinkurl{#1}}}
\providecommand{\doeprint}[1]{\href{http://ascl.net/#1}{\nolinkurl{http://ascl.net/#1}}}
\providecommand{\doarXiv}[1]{\href{https://arxiv.org/abs/#1}{\nolinkurl{https://arxiv.org/abs/#1}}}

\bibitem[{{Acciari} {et~al.}(2008){Acciari}, {Aliu}, {Beilicke}, {Benbow},
  {B{\"o}ttcher}, {Bradbury}, {Buckley}, {Bugaev}, {Butt}, {Celik}, {Cesarini},
  {Ciupik}, {Chow}, {Cogan}, {Colin}, {Cui}, {Daniel}, {Ergin}, {Falcone},
  {Fegan}, {Finley}, {Finnegan}, {Fortin}, {Fortson}, {Furniss}, {Gall},
  {Gillanders}, {Grube}, {Guenette}, {Gyuk}, {Hanna}, {Hays}, {Holder},
  {Horan}, {Hui}, {Humensky}, {Imran}, {Kaaret}, {Karlsson}, {Kertzman},
  {Kieda}, {Konopelko}, {Krawczynski}, {Krennrich}, {Lang}, {LeBohec}, {Lee},
  {Maier}, {McCann}, {McCutcheon}, {Moriarty}, {Mukherjee}, {Nagai}, {Niemiec},
  {Ong}, {Pandel}, {Perkins}, {Petry}, {Pohl}, {Quinn}, {Ragan}, {Reyes},
  {Reynolds}, {Roache}, {Rose}, {Schroedter}, {Sembroski}, {Smith}, {Steele},
  {Swordy}, {Toner}, {Vassiliev}, {Wagner}, {Wakely}, {Ward}, {Weekes},
  {Weinstein}, {White}, {Williams}, {Wissel}, {Wood}, \&
  {Zitzer}}]{2008ApJ...684L..73A}
{Acciari}, V.~A., {Aliu}, E., {Beilicke}, M., {et~al.} 2008, \apjl, 684, L73

\bibitem[{{Acciari} {et~al.}(2009){Acciari}, {Aliu}, {Aune}, {Beilicke},
  {Benbow}, {B{\"o}ttcher}, {Boltuch}, {Buckley}, {Bradbury}, {Bugaev},
  {Byrum}, {Cannon}, {Cesarini}, {Ciupik}, {Cogan}, {Cui}, {Dickherber},
  {Duke}, {Falcone}, {Finley}, {Fortin}, {Fortson}, {Furniss}, {Galante},
  {Gall}, {Gibbs}, {Gillanders}, {Grube}, {Guenette}, {Gyuk}, {Hanna},
  {Holder}, {Hui}, {Humensky}, {Kaaret}, {Karlsson}, {Kertzman}, {Kieda},
  {Konopelko}, {Krawczynski}, {Krennrich}, {Lang}, {Le Bohec}, {Maier},
  {McArthur}, {McCann}, {McCutcheon}, {Millis}, {Moriarty}, {Ong}, {Otte},
  {Pandel}, {Perkins}, {Pichel}, {Pohl}, {Quinn}, {Ragan}, {Reyes}, {Reynolds},
  {Roache}, {Rose}, {Sembroski}, {Smith}, {Steele}, {Theiling}, {Thibadeau},
  {Varlotta}, {Vassiliev}, {Vincent}, {Wakely}, {Ward}, {Weekes}, {Weinstein},
  {Weisgarber}, {Williams}, {Wissel}, {Wood}, {Pian}, {Vercellone},
  {Donnarumma}, {D'Ammando}, {Bulgarelli}, {Chen}, {Giuliani}, {Longo},
  {Pacciani}, {Pucella}, {Vittorini}, {Tavani}, {Argan}, {Barbiellini},
  {Caraveo}, {Cattaneo}, {Cocco}, {Costa}, {Del Monte}, {De Paris}, {Di Cocco},
  {Evangelista}, {Feroci}, {Fiorini}, {Froysland}, {Frutti}, {Fuschino},
  {Galli}, {Gianotti}, {Labanti}, {Lapshov}, {Lazzarotto}, {Lipari},
  {Marisaldi}, {Mastropietro}, {Mereghetti}, {Morelli}, {Morselli},
  {Pellizzoni}, {Perotti}, {Piano}, {Picozza}, {Pilia}, {Porrovecchio},
  {Prest}, {Rapisarda}, {Rappoldi}, {Rubini}, {Sabatini}, {Soffitta},
  {Trifoglio}, {Trois}, {Vallazza}, {Zambra}, {Zanello}, {Pittori},
  {Santolamazza}, {Verrecchia}, {Giommi}, {Colafrancesco}, {Salotti},
  {Villata}, {Raiteri}, {Aller}, {Aller}, {Arkharov}, {Efimova}, {Larionov},
  {Leto}, {Ligustri}, {Lindfors}, {Pasanen}, {Kurtanidze}, {Tetradze},
  {Lahteenmaki}, {Kotiranta}, {Cucchiara}, {Romano}, {Nesci}, {Pursimo},
  {Heidt}, {Benitez}, {Hiriart}, {Nilsson}, {Berdyugin}, {Mujica}, {Dultzin},
  {Lopez}, {Mommert}, {Sorcia}, \& {de la Calle Perez}}]{2009ApJ...707..612A}
{Acciari}, V.~A., {Aliu}, E., {Aune}, T., {et~al.} 2009, \apj, 707, 612

\bibitem[{{Acero} {et~al.}(2015){Acero}, {Ackermann}, {Ajello}, {Albert},
  {Atwood}, {Axelsson}, {Baldini}, {Ballet}, {Barbiellini}, {Bastieri},
  {Belfiore}, {Bellazzini}, {Bissaldi}, {Blandford}, {Bloom}, {Bogart},
  {Bonino}, {Bottacini}, {Bregeon}, {Britto}, {Bruel}, {Buehler}, {Burnett},
  {Buson}, {Caliandro}, {Cameron}, {Caputo}, {Caragiulo}, {Caraveo},
  {Casandjian}, {Cavazzuti}, {Charles}, {Chaves}, {Chekhtman}, {Cheung},
  {Chiang}, {Chiaro}, {Ciprini}, {Claus}, {Cohen-Tanugi}, {Cominsky}, {Conrad},
  {Cutini}, {D'Ammando}, {de Angelis}, {DeKlotz}, {de Palma}, {Desiante},
  {Digel}, {Di Venere}, {Drell}, {Dubois}, {Dumora}, {Favuzzi}, {Fegan},
  {Ferrara}, {Finke}, {Franckowiak}, {Fukazawa}, {Funk}, {Fusco}, {Gargano},
  {Gasparrini}, {Giebels}, {Giglietto}, {Giommi}, {Giordano}, {Giroletti},
  {Glanzman}, {Godfrey}, {Grenier}, {Grondin}, {Grove}, {Guillemot}, {Guiriec},
  {Hadasch}, {Harding}, {Hays}, {Hewitt}, {Hill}, {Horan}, {Iafrate}, {Jogler},
  {J{\'o}hannesson}, {Johnson}, {Johnson}, {Johnson}, {Johnson}, {Kamae},
  {Kataoka}, {Katsuta}, {Kuss}, {La Mura}, {Landriu}, {Larsson}, {Latronico},
  {Lemoine-Goumard}, {Li}, {Li}, {Longo}, {Loparco}, {Lott}, {Lovellette},
  {Lubrano}, {Madejski}, {Massaro}, {Mayer}, {Mazziotta}, {McEnery},
  {Michelson}, {Mirabal}, {Mizuno}, {Moiseev}, {Mongelli}, {Monzani},
  {Morselli}, {Moskalenko}, {Murgia}, {Nuss}, {Ohno}, {Ohsugi}, {Omodei},
  {Orienti}, {Orlando}, {Ormes}, {Paneque}, {Panetta}, {Perkins},
  {Pesce-Rollins}, {Piron}, {Pivato}, {Porter}, {Racusin}, {Rando}, {Razzano},
  {Razzaque}, {Reimer}, {Reimer}, {Reposeur}, {Rochester}, {Romani},
  {Salvetti}, {S{\'a}nchez-Conde}, {Saz Parkinson}, {Schulz}, {Siskind},
  {Smith}, {Spada}, {Spandre}, {Spinelli}, {Stephens}, {Strong}, {Suson},
  {Takahashi}, {Takahashi}, {Tanaka}, {Thayer}, {Thayer}, {Thompson},
  {Tibaldo}, {Tibolla}, {Torres}, {Torresi}, {Tosti}, {Troja}, {Van Klaveren},
  {Vianello}, {Winer}, {Wood}, {Wood}, {Zimmer}, \& {Fermi-LAT
  Collaboration}}]{3FGL}
{Acero}, F., {Ackermann}, M., {Ajello}, M., {et~al.} 2015, \apjs, 218, 23

\bibitem[{{Ackermann} {et~al.}(2011){Ackermann}, {Ajello}, {Allafort},
  {Antolini}, {Atwood}, {Axelsson}, {Baldini}, {Ballet}, {Barbiellini},
  {Bastieri}, {Bechtol}, {Bellazzini}, {Berenji}, {Blandford}, {Bloom},
  {Bonamente}, {Borgland}, {Bottacini}, {Bouvier}, {Bregeon}, {Brigida},
  {Bruel}, {Buehler}, {Burnett}, {Buson}, {Caliandro}, {Cameron}, {Caraveo},
  {Casandjian}, {Cavazzuti}, {Cecchi}, {Charles}, {Cheung}, {Chiang},
  {Ciprini}, {Claus}, {Cohen-Tanugi}, {Conrad}, {Costamante}, {Cutini}, {de
  Angelis}, {de Palma}, {Dermer}, {Digel}, {Silva}, {Drell}, {Dubois},
  {Escande}, {Favuzzi}, {Fegan}, {Ferrara}, {Finke}, {Focke}, {Fortin},
  {Frailis}, {Fukazawa}, {Funk}, {Fusco}, {Gargano}, {Gasparrini}, {Gehrels},
  {Germani}, {Giebels}, {Giglietto}, {Giommi}, {Giordano}, {Giroletti},
  {Glanzman}, {Godfrey}, {Grenier}, {Grove}, {Guiriec}, {Gustafsson},
  {Hadasch}, {Hayashida}, {Hays}, {Healey}, {Horan}, {Hou}, {Hughes},
  {Iafrate}, {J{\'o}hannesson}, {Johnson}, {Johnson}, {Kamae}, {Katagiri},
  {Kataoka}, {Kn{\"o}dlseder}, {Kuss}, {Lande}, {Larsson}, {Latronico},
  {Longo}, {Loparco}, {Lott}, {Lovellette}, {Lubrano}, {Madejski}, {Mazziotta},
  {McConville}, {McEnery}, {Michelson}, {Mitthumsiri}, {Mizuno}, {Moiseev},
  {Monte}, {Monzani}, {Moretti}, {Morselli}, {Moskalenko}, {Murgia},
  {Nakamori}, {Naumann-Godo}, {Nolan}, {Norris}, {Nuss}, {Ohno}, {Ohsugi},
  {Okumura}, {Omodei}, {Orienti}, {Orlando}, {Ormes}, {Ozaki}, {Paneque},
  {Parent}, {Pesce-Rollins}, {Pierbattista}, {Piranomonte}, {Piron}, {Pivato},
  {Porter}, {Rain{\`o}}, {Rando}, {Razzano}, {Razzaque}, {Reimer}, {Reimer},
  {Ritz}, {Rochester}, {Romani}, {Roth}, {Sanchez}, {Sbarra}, {Scargle},
  {Schalk}, {Sgr{\`o}}, {Shaw}, {Siskind}, {Spandre}, {Spinelli}, {Strong},
  {Suson}, {Tajima}, {Takahashi}, {Takahashi}, {Tanaka}, {Thayer}, {Thayer},
  {Thompson}, {Tibaldo}, {Tinivella}, {Torres}, {Tosti}, {Troja}, {Uchiyama},
  {Vandenbroucke}, {Vasileiou}, {Vianello}, {Vitale}, {Waite}, {Wallace},
  {Wang}, {Winer}, {Wood}, {Wood}, \& {Zimmer}}]{2LAC}
{Ackermann}, M., {Ajello}, M., {Allafort}, A., {et~al.} 2011, \apj, 743, 171

\bibitem[{{Ackermann} {et~al.}(2012){Ackermann}, {Ajello}, {Allafort},
  {Schady}, {Baldini}, {Ballet}, {Barbiellini}, {Bastieri}, {Bellazzini},
  {Blandford}, {Bloom}, {Borgland}, {Bottacini}, {Bouvier}, {Bregeon},
  {Brigida}, {Bruel}, {Buehler}, {Buson}, {Caliandro}, {Cameron}, {Caraveo},
  {Cavazzuti}, {Cecchi}, {Charles}, {Chaves}, {Chekhtman}, {Cheung}, {Chiang},
  {Chiaro}, {Ciprini}, {Claus}, {Cohen-Tanugi}, {Conrad}, {Cutini},
  {D'Ammando}, {de Palma}, {Dermer}, {Digel}, {do Couto e Silva},
  {Dom{\'{\i}}nguez}, {Drell}, {Drlica-Wagner}, {Favuzzi}, {Fegan}, {Focke},
  {Franckowiak}, {Fukazawa}, {Funk}, {Fusco}, {Gargano}, {Gasparrini},
  {Gehrels}, {Germani}, {Giglietto}, {Giordano}, {Giroletti}, {Glanzman},
  {Godfrey}, {Grenier}, {Grove}, {Guiriec}, {Gustafsson}, {Hadasch},
  {Hayashida}, {Hays}, {Jackson}, {Jogler}, {Kataoka}, {Kn{\"o}dlseder},
  {Kuss}, {Lande}, {Larsson}, {Latronico}, {Longo}, {Loparco}, {Lovellette},
  {Lubrano}, {Mazziotta}, {McEnery}, {Mehault}, {Michelson}, {Mizuno}, {Monte},
  {Monzani}, {Morselli}, {Moskalenko}, {Murgia}, {Tramacere}, {Nuss},
  {Greiner}, {Ohno}, {Ohsugi}, {Omodei}, {Orienti}, {Orlando}, {Ormes},
  {Paneque}, {Perkins}, {Pesce-Rollins}, {Piron}, {Pivato}, {Porter},
  {Rain{\`o}}, {Rando}, {Razzano}, {Razzaque}, {Reimer}, {Reimer}, {Reyes},
  {Ritz}, {Rau}, {Romoli}, {Roth}, {S{\'a}nchez-Conde}, {Sanchez}, {Scargle},
  {Sgr{\`o}}, {Siskind}, {Spandre}, {Spinelli}, {Stawarz}, {Suson},
  {Takahashi}, {Tanaka}, {Thayer}, {Thompson}, {Tibaldo}, {Tinivella},
  {Torres}, {Tosti}, {Troja}, {Usher}, {Vandenbroucke}, {Vasileiou},
  {Vianello}, {Vitale}, {Waite}, {Winer}, {Wood}, \& {Wood}}]{Fermi_EBL2012}
---. 2012, \sci, 338, 1190

\bibitem[{{Ackermann} {et~al.}(2013){Ackermann}, {Ajello}, {Allafort}, {Asano},
  {Atwood}, {Baldini}, {Ballet}, {Barbiellini}, {Bastieri}, {Bechtol},
  {Bellazzini}, {Bloom}, {Bonamente}, {Borgland}, {Bottacini}, {Brandt},
  {Bregeon}, {Brigida}, {Bruel}, {Buehler}, {Burnett}, {Busetto}, {Buson},
  {Caliandro}, {Cameron}, {Caraveo}, {Casandjian}, {Cecchi}, {Charles},
  {Chaty}, {Chekhtman}, {Cheung}, {Chiang}, {Cillis}, {Ciprini}, {Claus},
  {Cohen-Tanugi}, {Colafrancesco}, {Conrad}, {Cutini}, {D'Ammando}, {de Palma},
  {Dermer}, {Silva}, {Drell}, {Drlica-Wagner}, {Dubois}, {Favuzzi}, {Fegan},
  {Ferrara}, {Focke}, {Fortin}, {Fukazawa}, {Funk}, {Fusco}, {Gargano},
  {Gasparrini}, {Gehrels}, {Germani}, {Giglietto}, {Giordano}, {Giroletti},
  {Glanzman}, {Godfrey}, {Grandi}, {Grenier}, {Grove}, {Guiriec}, {Hadasch},
  {Hayashida}, {Hays}, {Horan}, {Hou}, {Hughes}, {Jackson}, {Jogler},
  {J{\'o}hannesson}, {Johnson}, {Johnson}, {Kamae}, {Kataoka}, {Kerr},
  {Kn{\"o}dlseder}, {Kuss}, {Lande}, {Larsson}, {Latronico}, {Lavalley}, {Lee},
  {Longo}, {Loparco}, {Lott}, {Lovellette}, {Lubrano}, {Mazziotta},
  {McConville}, {McEnery}, {Mehault}, {Michelson}, {Mignani}, {Mitthumsiri},
  {Mizuno}, {Moiseev}, {Monte}, {Monzani}, {Morselli}, {Moskalenko}, {Murgia},
  {Naumann-Godo}, {Nemmen}, {Nishino}, {Norris}, {Nuss}, {Ohsugi}, {Omodei},
  {Orienti}, {Orlando}, {Ormes}, {Paneque}, {Panetta}, {Pelassa}, {Perkins},
  {Pesce-Rollins}, {Pierbattista}, {Piron}, {Pivato}, {Poon}, {Porter},
  {Rain{\`o}}, {Rando}, {Razzano}, {Razzaque}, {Reimer}, {Reimer}, {Reyes},
  {Ritz}, {Rochester}, {Romoli}, {Roth}, {Sanchez}, {Saz Parkinson}, {Scargle},
  {Sgr{\`o}}, {Siskind}, {Snyder}, {Spandre}, {Spinelli}, {Stephens}, {Suson},
  {Tajima}, {Takahashi}, {Tanaka}, {Thayer}, {Thayer}, {Thompson}, {Tibaldo},
  {Tibolla}, {Tinivella}, {Tosti}, {Troja}, {Usher}, {Vandenbroucke},
  {Vasileiou}, {Vianello}, {Vitale}, {von Kienlin}, {Waite}, {Wallace},
  {Weltevrede}, {Winer}, {Wood}, {Wood}, {Yang}, \& {Zimmer}}]{FLAT-stack:2013}
---. 2013, \apj, 765, 54

\bibitem[{{Ackermann} {et~al.}(2015){Ackermann}, {Ajello}, {Atwood}, {Baldini},
  {Ballet}, {Barbiellini}, {Bastieri}, {Becerra Gonzalez}, {Bellazzini},
  {Bissaldi}, {Blandford}, {Bloom}, {Bonino}, {Bottacini}, {Brandt}, {Bregeon},
  {Britto}, {Bruel}, {Buehler}, {Buson}, {Caliandro}, {Cameron}, {Caragiulo},
  {Caraveo}, {Carpenter}, {Casandjian}, {Cavazzuti}, {Cecchi}, {Charles},
  {Chekhtman}, {Cheung}, {Chiang}, {Chiaro}, {Ciprini}, {Claus},
  {Cohen-Tanugi}, {Cominsky}, {Conrad}, {Cutini}, {D'Abrusco}, {D'Ammando}, {de
  Angelis}, {Desiante}, {Digel}, {Di Venere}, {Drell}, {Favuzzi}, {Fegan},
  {Ferrara}, {Finke}, {Focke}, {Franckowiak}, {Fuhrmann}, {Fukazawa},
  {Furniss}, {Fusco}, {Gargano}, {Gasparrini}, {Giglietto}, {Giommi},
  {Giordano}, {Giroletti}, {Glanzman}, {Godfrey}, {Grenier}, {Grove},
  {Guiriec}, {Hewitt}, {Hill}, {Horan}, {Itoh}, {J{\'o}hannesson}, {Johnson},
  {Johnson}, {Kataoka}, {Kawano}, {Krauss}, {Kuss}, {La Mura}, {Larsson},
  {Latronico}, {Leto}, {Li}, {Li}, {Longo}, {Loparco}, {Lott}, {Lovellette},
  {Lubrano}, {Madejski}, {Mayer}, {Mazziotta}, {McEnery}, {Michelson},
  {Mizuno}, {Moiseev}, {Monzani}, {Morselli}, {Moskalenko}, {Murgia}, {Nuss},
  {Ohno}, {Ohsugi}, {Ojha}, {Omodei}, {Orienti}, {Orlando}, {Paggi}, {Paneque},
  {Perkins}, {Pesce-Rollins}, {Piron}, {Pivato}, {Porter}, {Rain{\`o}},
  {Rando}, {Razzano}, {Razzaque}, {Reimer}, {Reimer}, {Romani}, {Salvetti},
  {Schaal}, {Schinzel}, {Schulz}, {Sgr{\`o}}, {Siskind}, {Sokolovsky}, {Spada},
  {Spandre}, {Spinelli}, {Stawarz}, {Suson}, {Takahashi}, {Takahashi},
  {Tanaka}, {Thayer}, {Thayer}, {Tibaldo}, {Torres}, {Torresi}, {Tosti},
  {Troja}, {Uchiyama}, {Vianello}, {Winer}, {Wood}, \& {Zimmer}}]{3LAC}
{Ackermann}, M., {Ajello}, M., {Atwood}, W.~B., {et~al.} 2015, \apj, 810, 14

\bibitem[{{Aharonian} {et~al.}(2006){Aharonian}, {Akhperjanian}, {Bazer-Bachi},
  {Beilicke}, {Benbow}, {Berge}, {Bernl{\"o}hr}, {Boisson}, {Bolz}, {Borrel},
  {Braun}, {Breitling}, {Brown}, {Chadwick}, {Chounet}, {Cornils},
  {Costamante}, {Degrange}, {Dickinson}, {Djannati-Ata{\"i}}, {Drury}, {Dubus},
  {Emmanoulopoulos}, {Espigat}, {Feinstein}, {Fontaine}, {Fuchs}, {Funk},
  {Gallant}, {Giebels}, {Gillessen}, {Glicenstein}, {Goret}, {Hadjichristidis},
  {Hauser}, {Hauser}, {Heinzelmann}, {Henri}, {Hermann}, {Hinton}, {Hofmann},
  {Holleran}, {Horns}, {Jacholkowska}, {de Jager}, {Kh{\'e}lifi}, {Klages},
  {Komin}, {Konopelko}, {Latham}, {Le Gallou}, {Lemi{\`e}re},
  {Lemoine-Goumard}, {Leroy}, {Lohse}, {Martin}, {Martineau-Huynh},
  {Marcowith}, {Masterson}, {McComb}, {de Naurois}, {Nolan}, {Noutsos},
  {Orford}, {Osborne}, {Ouchrif}, {Panter}, {Pelletier}, {Pita},
  {P{\"u}hlhofer}, {Punch}, {Raubenheimer}, {Raue}, {Raux}, {Rayner}, {Reimer},
  {Reimer}, {Ripken}, {Rob}, {Rolland}, {Rowell}, {Sahakian}, {Saug{\'e}},
  {Schlenker}, {Schlickeiser}, {Schuster}, {Schwanke}, {Siewert}, {Sol},
  {Spangler}, {Steenkamp}, {Stegmann}, {Tavernet}, {Terrier}, {Th{\'e}oret},
  {Tluczykont}, {van Eldik}, {Vasileiadis}, {Venter}, {Vincent}, {V{\"o}lk}, \&
  {Wagner}}]{Ahar_etal:06}
{Aharonian}, F., {Akhperjanian}, A.~G., {Bazer-Bachi}, A.~R., {et~al.} 2006,
  \nat, 440, 1018

\bibitem[{{Aharonian} {et~al.}(1994){Aharonian}, {Coppi}, \&
  {Voelk}}]{1994ApJ...423L...5A}
{Aharonian}, F.~A., {Coppi}, P.~S., \& {Voelk}, H.~J. 1994, \apjl, 423, L5

\bibitem[{{Ahnen} {et~al.}(2018){Ahnen}, {Ansoldi}, {Antonelli}, {Arcaro},
  {Baack}, {Babi{\'c}}, {Banerjee}, {Bangale}, {Barres de Almeida}, {Abel
  Barrio}, {Becerra Gonz{\'a}lez}, {Bednarek}, {Bernardini}, {Berti},
  {Bhattacharyya}, {Biland}, {Blanch}, {Bonnoli}, {Carosi}, {Carosi},
  {Ceribella}, {Chatterjee}, {Merve Colak}, {Colin}, {Colombo}, {Contreras},
  {Cortina}, {Covino}, {Cumani}, {Da Vela}, {Dazzi}, {De Angelis}, {De Lotto},
  {Delfino}, {Delgado}, {Di Pierro}, {Dom{\'{\i}}nguez}, {Dominis Prester},
  {Dorner}, {Doro}, {Einecke}, {Elsaesser}, {Fallah Ramazani},
  {Fern{\'a}ndez-Barral}, {Fidalgo}, {Foffano}, {Pfrang}, {Fonseca}, {Font},
  {Fruck}, {Galindo}, {Gallozzi}, {Garc{\'{\i}}a L{\'o}pez}, {Garczarczyk},
  {Gaug}, {Giammaria}, {Godinovi{\'c}}, {Gora}, {Guberman}, {Hadasch}, {Hahn},
  {Hassan}, {Hayashida}, {Herrera}, {Hose}, {Hrupec}, {Inoue}, {Ishio},
  {Iwamura}, {Konno}, {Kubo}, {Kushida}, {Lelas}, {Lindfors}, {Lombardi},
  {Longo}, {L{\'o}pez}, {Maggio}, {Majumdar}, {Makariev}, {Maneva},
  {Manganaro}, {Mannheim}, {Maraschi}, {Mariotti}, {Mart{\'{\i}}nez}, {Masuda},
  {Mazin}, {Minev}, {Miranda}, {Mirzoyan}, {Moralejo}, {Moreno}, {Moretti},
  {Nagayoshi}, {Neustroev}, {Niedzwiecki}, {Nievas Rosillo}, {Nigro},
  {Nilsson}, {Ninci}, {Nishijima}, {Noda}, {Nogu{\'e}s}, {Paiano}, {Palacio},
  {Paneque}, {Paoletti}, {Paredes}, {Pedaletti}, {Peresano}, {Persic}, {Prada
  Moroni}, {Prandini}, {Puljak}, {Rodriguez Garcia}, {Reichardt}, {Rhode},
  {Rib{\'o}}, {Rico}, {Righi}, {Rugliancich}, {Saito}, {Satalecka},
  {Schweizer}, {Sitarek}, {Snidari{\'c}}, {Sobczynska}, {Stamerra}, {Strzys},
  {Suri{\'c}}, {Takahashi}, {Tavecchio}, {Temnikov}, {Terzi{\'c}}, {Teshima},
  {Torres-Alb{\`a}}, {Treves}, {Tsujimoto}, {Vanzo}, {Vazquez Acosta}, {Vovk},
  {Ward}, {Will}, {Zari{\'c}}, \& {Cerruti}}]{Ahnen2018}
{Ahnen}, M.~L., {Ansoldi}, S., {Antonelli}, L.~A., {et~al.} 2018,
  arXiv:1807.04300

\bibitem[{{Albert} {et~al.}(2007){Albert}, {Aliu}, {Anderhub}, {Antoranz},
  {Armada}, {Asensio}, {Baixeras}, {Barrio}, {Bartko}, {Bastieri}, {Becker},
  {Bednarek}, {Berger}, {Bigongiari}, {Biland}, {Bock}, {Bordas},
  {Bosch-Ramon}, {Bretz}, {Britvitch}, {Camara}, {Carmona}, {Chilingarian},
  {Ciprini}, {Coarasa}, {Commichau}, {Contreras}, {Cortina}, {Curtef},
  {Danielyan}, {Dazzi}, {De Angelis}, {de los Reyes}, {De Lotto},
  {Domingo-Santamar{\'{\i}}a}, {Dorner}, {Doro}, {Errando}, {Fagiolini},
  {Ferenc}, {Fern{\'a}ndez}, {Firpo}, {Flix}, {Fonseca}, {Font}, {Fuchs},
  {Galante}, {Garczarczyk}, {Gaug}, {Giller}, {Goebel}, {Hakobyan},
  {Hayashida}, {Hengstebeck}, {H{\"o}hne}, {Hose}, {Hsu}, {Jacon}, {Jogler},
  {Kalekin}, {Kosyra}, {Kranich}, {Kritzer}, {Laatiaoui}, {Laille}, {Liebing},
  {Lindfors}, {Lombardi}, {Longo}, {L{\'o}pez}, {L{\'o}pez}, {Lorenz},
  {Majumdar}, {Maneva}, {Mannheim}, {Mansutti}, {Mariotti}, {Mart{\'{\i}}nez},
  {Mazin}, {Merck}, {Meucci}, {Meyer}, {Miranda}, {Mirzoyan}, {Mizobuchi},
  {Moralejo}, {Nilsson}, {Ninkovic}, {O{\~n}a-Wilhelmi}, {Ordu{\~n}a}, {Otte},
  {Oya}, {Paneque}, {Paoletti}, {Paredes}, {Pasanen}, {Pascoli}, {Pauss},
  {Pegna}, {Persic}, {Peruzzo}, {Piccioli}, {Poller}, {Prandini}, {Raymers},
  {Rhode}, {Rib{\'o}}, {Rico}, {Rissi}, {Robert}, {R{\"u}gamer}, {Saggion},
  {S{\'a}nchez}, {Sartori}, {Scalzotto}, {Scapin}, {Schmitt}, {Schweizer},
  {Shayduk}, {Shinozaki}, {Shore}, {Sidro}, {Sillanp{\"a}{\"a}}, {Sobczynska},
  {Stamerra}, {Stark}, {Takalo}, {Temnikov}, {Tescaro}, {Teshima}, {Tonello},
  {Torres}, {Torres}, {Turini}, {Vankov}, {Vitale}, {Wagner}, {Wibig},
  {Wittek}, {Zanin}, \& {Zapatero}}]{2007ApJ...663..125A}
{Albert}, J., {Aliu}, E., {Anderhub}, H., {et~al.} 2007, \apj, 663, 125

\bibitem[{{Aleksi{\'c}} {et~al.}(2011{\natexlab{a}}){Aleksi{\'c}}, {Antonelli},
  {Antoranz}, {Backes}, {Barrio}, {Bastieri}, {Becerra Gonz{\'a}lez},
  {Bednarek}, {Berdyugin}, {Berger}, {Bernardini}, {Biland}, {Blanch}, {Bock},
  {Boller}, {Bonnoli}, {Borla Tridon}, {Braun}, {Bretz}, {Ca{\~n}ellas},
  {Carmona}, {Carosi}, {Colin}, {Colombo}, {Contreras}, {Cortina}, {Cossio},
  {Covino}, {Dazzi}, {De Angelis}, {De Cea del Pozo}, {De Lotto}, {Delgado
  Mendez}, {Diago Ortega}, {Doert}, {Dom{\'{\i}}nguez}, {Dominis Prester},
  {Dorner}, {Doro}, {Elsaesser}, {Ferenc}, {Fonseca}, {Font}, {Fruck},
  {Garc{\'{\i}}a L{\'o}pez}, {Garczarczyk}, {Garrido}, {Giavitto},
  {Godinovi{\'c}}, {Hadasch}, {H{\"a}fner}, {Herrero}, {Hildebrand},
  {H{\"o}hne-M{\"o}nch}, {Hose}, {Hrupec}, {Huber}, {Jogler}, {Klepser},
  {Kr{\"a}henb{\"u}hl}, {Krause}, {La Barbera}, {Lelas}, {Leonardo},
  {Lindfors}, {Lombardi}, {L{\'o}pez}, {Lorenz}, {Makariev}, {Maneva},
  {Mankuzhiyil}, {Mannheim}, {Maraschi}, {Mariotti}, {Mart{\'{\i}}nez},
  {Mazin}, {Meucci}, {Miranda}, {Mirzoyan}, {Miyamoto}, {Mold{\'o}n},
  {Moralejo}, {Nieto}, {Nilsson}, {Orito}, {Oya}, {Paneque}, {Paoletti},
  {Pardo}, {Paredes}, {Partini}, {Pasanen}, {Pauss}, {Perez-Torres}, {Persic},
  {Peruzzo}, {Pilia}, {Pochon}, {Prada}, {Prada Moroni}, {Prandini}, {Puljak},
  {Reichardt}, {Reinthal}, {Rhode}, {Rib{\'o}}, {Rico}, {R{\"u}gamer},
  {Saggion}, {Saito}, {Saito}, {Salvati}, {Satalecka}, {Scalzotto}, {Scapin},
  {Schultz}, {Schweizer}, {Shayduk}, {Shore}, {Sillanp{\"a}{\"a}}, {Sitarek},
  {Sobczynska}, {Spanier}, {Spiro}, {Stamerra}, {Steinke}, {Storz}, {Strah},
  {Suri{\'c}}, {Takalo}, {Tavecchio}, {Temnikov}, {Terzi{\'c}}, {Tescaro},
  {Teshima}, {Thom}, {Tibolla}, {Torres}, {Treves}, {Vankov}, {Vogler},
  {Wagner}, {Weitzel}, {Zabalza}, {Zandanel}, {Zanin}, {MAGIC Collaboration},
  {Tanaka}, {Wood}, \& {Buson}}]{MAGIC_4c21_2010}
{Aleksi{\'c}}, J., {Antonelli}, L.~A., {Antoranz}, P., {et~al.}
  2011{\natexlab{a}}, \apjl, 730, L8

\bibitem[{{Aleksi{\'c}} {et~al.}(2011{\natexlab{b}}){Aleksi{\'c}}, {Antonelli},
  {Antoranz}, {Backes}, {Barrio}, {Bastieri}, {Becerra Gonz{\'a}lez},
  {Bednarek}, {Berdyugin}, {Berger}, {Bernardini}, {Biland}, {Blanch}, {Bock},
  {Boller}, {Bonnoli}, {Borla Tridon}, {Braun}, {Bretz}, {Ca{\~n}ellas},
  {Carmona}, {Carosi}, {Colin}, {Colombo}, {Contreras}, {Cortina}, {Cossio},
  {Covino}, {Dazzi}, {de Angelis}, {de Cea Del Pozo}, {de Lotto}, {Delgado
  Mendez}, {Diago Ortega}, {Doert}, {Dom{\'{\i}}nguez}, {Dominis Prester},
  {Dorner}, {Doro}, {Elsaesser}, {Ferenc}, {Fonseca}, {Font}, {Fruck},
  {Garc{\'{\i}}a L{\'o}pez}, {Garczarczyk}, {Garrido}, {Giavitto},
  {Godinovi{\'c}}, {Hadasch}, {H{\"a}fner}, {Herrero}, {Hildebrand}, {Hose},
  {Hrupec}, {Huber}, {Jogler}, {Klepser}, {Kr{\"a}henb{\"u}hl}, {Krause}, {La
  Barbera}, {Lelas}, {Leonardo}, {Lindfors}, {Lombardi}, {L{\'o}pez}, {Lorenz},
  {Majumdar}, {Makariev}, {Maneva}, {Mankuzhiyil}, {Mannheim}, {Maraschi},
  {Mariotti}, {Mart{\'{\i}}nez}, {Mazin}, {Meucci}, {Miranda}, {Mirzoyan},
  {Miyamoto}, {Mold{\'o}n}, {Moralejo}, {Nieto}, {Nilsson}, {Orito}, {Oya},
  {Paoletti}, {Pardo}, {Paredes}, {Partini}, {Pasanen}, {Pauss},
  {Perez-Torres}, {Persic}, {Peruzzo}, {Pilia}, {Pochon}, {Prada}, {Prada
  Moroni}, {Prandini}, {Puljak}, {Reichardt}, {Reinthal}, {Rhode}, {Rib{\'o}},
  {Rico}, {R{\"u}gamer}, {R{\"u}ger}, {Saggion}, {Saito}, {Saito}, {Salvati},
  {Satalecka}, {Scalzotto}, {Scapin}, {Schultz}, {Schweizer}, {Shayduk},
  {Shore}, {Sillanp{\"a}{\"a}}, {Sitarek}, {Sobczynska}, {Spanier}, {Spiro},
  {Stamerra}, {Steinke}, {Storz}, {Strah}, {Suri{\'c}}, {Takalo}, {Tavecchio},
  {Temnikov}, {Terzi{\'c}}, {Tescaro}, {Teshima}, {Thom}, {Tibolla}, {Torres},
  {Treves}, {Vankov}, {Vogler}, {Wagner}, {Weitzel}, {Zabalza}, {Zandanel}, \&
  {Zanin}}]{MAGIC_3c279_2007}
---. 2011{\natexlab{b}}, \aap, 530, A4

\bibitem[{{Allevato} {et~al.}(2014){Allevato}, {Finoguenov}, \&
  {Cappelluti}}]{Allevato2014}
{Allevato}, V., {Finoguenov}, A., \& {Cappelluti}, N. 2014, \apj, 797, 96

\bibitem[{{Ando} \& {Kusenko}(2010)}]{Ando:2010}
{Ando}, S., \& {Kusenko}, A. 2010, \apjl, 722, L39

\bibitem[{{B{\^i}rzan} {et~al.}(2008){B{\^i}rzan}, {McNamara}, {Nulsen},
  {Carilli}, \& {Wise}}]{Birzan08}
{B{\^i}rzan}, L., {McNamara}, B.~R., {Nulsen}, P.~E.~J., {Carilli}, C.~L., \&
  {Wise}, M.~W. 2008, \apj, 686, 859

\bibitem[{{Blumenthal} \& {Gould}(1970)}]{BlumenthalGould1970}
{Blumenthal}, G.~R., \& {Gould}, R.~J. 1970, Reviews of Modern Physics, 42, 237

\bibitem[{{Broderick} {et~al.}(2012){Broderick}, {Chang}, \&
  {Pfrommer}}]{PaperI}
{Broderick}, A.~E., {Chang}, P., \& {Pfrommer}, C. 2012, \apj, 752, 22

\bibitem[{{Broderick} {et~al.}(2016){Broderick}, {Tiede}, {Shalaby},
  {Pfrommer}, {Puchwein}, {Chang}, \& {Lamberts}}]{BTI}
{Broderick}, A.~E., {Tiede}, P., {Shalaby}, M., {et~al.} 2016, \apj, 832, 109

\bibitem[{{Capetti} {et~al.}(2017){Capetti}, {Massaro}, \& {Baldi}}]{FRICAT}
{Capetti}, A., {Massaro}, F., \& {Baldi}, R.~D. 2017, \aap, 598, A49

\bibitem[{{Cerruti} {et~al.}(2018){Cerruti}, {Zech}, {Boisson}, {Emery},
  {Inoue}, \& {Lenain}}]{Cerruti2018}
{Cerruti}, M., {Zech}, A., {Boisson}, C., {et~al.} 2018, arXiv:1807.04335

\bibitem[{{Cerruti} {et~al.}(2015)}]{VERITAS_4c21_2014b}
{Cerruti}, M., {et~al.} 2015, arXiv:1501.03554

\bibitem[{{Chang} {et~al.}(2012){Chang}, {Broderick}, \& {Pfrommer}}]{PaperII}
{Chang}, P., {Broderick}, A.~E., \& {Pfrommer}, C. 2012, \apj, 752, 23

\bibitem[{{Chang} {et~al.}(2014){Chang}, {Broderick}, {Pfrommer}, {Puchwein},
  {Lamberts}, \& {Shalaby}}]{Chang:2014}
{Chang}, P., {Broderick}, A.~E., {Pfrommer}, C., {et~al.} 2014, \apj, 797, 110

\bibitem[{{Dermer} {et~al.}(2011){Dermer}, {Cavadini}, {Razzaque}, {Finke},
  {Chiang}, \& {Lott}}]{Derm_etal:11}
{Dermer}, C.~D., {Cavadini}, M., {Razzaque}, S., {et~al.} 2011, \apjl, 733, L21

\bibitem[{{Dom{\'{\i}}nguez} {et~al.}(2011){Dom{\'{\i}}nguez}, {Primack},
  {Rosario}, {Prada}, {Gilmore}, {Faber}, {Koo}, {Somerville},
  {P{\'e}rez-Torres}, {P{\'e}rez-Gonz{\'a}lez}, {Huang}, {Davis},
  {Guhathakurta}, {Barmby}, {Conselice}, {Lozano}, {Newman}, \&
  {Cooper}}]{Dominguez11}
{Dom{\'{\i}}nguez}, A., {Primack}, J.~R., {Rosario}, D.~J., {et~al.} 2011,
  \mnras, 410, 2556

\bibitem[{{Duplessis} \& {Vachaspati}(2017)}]{2017JCAP...05..005D}
{Duplessis}, F., \& {Vachaspati}, T. 2017, \jcap, 5, 005

\bibitem[{{Elyiv} {et~al.}(2009){Elyiv}, {Neronov}, \&
  {Semikoz}}]{2009PhRvD..80b3010E}
{Elyiv}, A., {Neronov}, A., \& {Semikoz}, D.~V. 2009, \prd, 80, 023010

\bibitem[{{Fermi-LAT Collaboration} \& {Biteau}(2018)}]{FermiIGMF:18}
{Fermi-LAT Collaboration}, \& {Biteau}, J. 2018, arXiv:1804.08035

\bibitem[{{Foreman-Mackey} {et~al.}(2013){Foreman-Mackey}, {Hogg}, {Lang}, \&
  {Goodman}}]{emcee}
{Foreman-Mackey}, D., {Hogg}, D.~W., {Lang}, D., \& {Goodman}, J. 2013, \pasp,
  125, 306

\bibitem[{{Gao} {et~al.}(2018){Gao}, {Fedynitch}, {Winter}, \&
  {Pohl}}]{Gao2018}
{Gao}, S., {Fedynitch}, A., {Winter}, W., \& {Pohl}, M. 2018, arXiv:1807.04275

\bibitem[{{Gould} \& {Schr{\'e}der}(1966)}]{Gould+66}
{Gould}, R.~J., \& {Schr{\'e}der}, G. 1966, Physical Review Letters, 16, 252

\bibitem[{Greisen(1966)}]{Greisen1966}
Greisen, K. 1966, Phys. Rev. Lett., 16, 748

\bibitem[{{Helfand} {et~al.}(2015){Helfand}, {White}, \& {Becker}}]{FIRST_CAT}
{Helfand}, D.~J., {White}, R.~L., \& {Becker}, R.~H. 2015, \apj, 801, 26

\bibitem[{{Holder}(2014)}]{VERITAS_4c21_2014}
{Holder}, J. 2014, The Astronomer's Telegram, 5981

\bibitem[{{Hopkins} {et~al.}(2007){Hopkins}, {Richards}, \&
  {Hernquist}}]{Hopkins+07}
{Hopkins}, P.~F., {Richards}, G.~T., \& {Hernquist}, L. 2007, \apj, 654, 731

\bibitem[{{IceCube Collaboration}(2018{\natexlab{a}})}]{TXSblazarI}
{IceCube Collaboration}. 2018{\natexlab{a}}, Science, 361

\bibitem[{{IceCube Collaboration}(2018{\natexlab{b}})}]{TXSblazarII}
---. 2018{\natexlab{b}}, Science

\bibitem[{{Kapi{\'n}ska} {et~al.}(2017){Kapi{\'n}ska}, {Terentev}, {Wong},
  {Shabala}, {Andernach}, {Rudnick}, {Storer}, {Banfield}, {Willett}, {de
  Gasperin}, {Lintott}, {L{\'o}pez-S{\'a}nchez}, {Middelberg}, {Norris},
  {Schawinski}, {Seymour}, \& {Simmons}}]{Kapinska+17}
{Kapi{\'n}ska}, A.~D., {Terentev}, I., {Wong}, O.~I., {et~al.} 2017, \aj, 154,
  253

\bibitem[{{Kharb} {et~al.}(2010){Kharb}, {Lister}, \& {Cooper}}]{Kharb+10}
{Kharb}, P., {Lister}, M.~L., \& {Cooper}, N.~J. 2010, \apj, 710, 764

\bibitem[{{Kulsrud} \& {Zweibel}(2008)}]{2008RPPh...71d6901K}
{Kulsrud}, R.~M., \& {Zweibel}, E.~G. 2008, Reports on Progress in Physics, 71,
  046901

\bibitem[{{Lamberts} {et~al.}(2015){Lamberts}, {Chang}, {Pfrommer}, {Puchwein},
  {Broderick}, \& {Shalaby}}]{Lamberts:2015}
{Lamberts}, A., {Chang}, P., {Pfrommer}, C., {et~al.} 2015, \apj, 811, 19

\bibitem[{{Landt} \& {Bignall}(2008)}]{Landt+08}
{Landt}, H., \& {Bignall}, H.~E. 2008, \mnras, 391, 967

\bibitem[{{Lindfors} {et~al.}(2013){Lindfors}, {Nilsson}, {Barres de Almeida},
  {Mazin}, {Paneque}, {Saito}, {Becerra Gonzalez}, {Berger}, {De Caneva},
  {Schultz}, {Sitarek}, {Stamerra}, {Tavecchio}, {on behalf of the MAGIC
  collaboration}, {Buson}, {D'Ammando}, {Hayashida}, {on behalf of the
  Fermi-LAT collaboration A.~L{\"a}hteenm{\"a}ki}, {Tornikoski}, \&
  {Hovatta}}]{MAGIC_3c279_2011}
{Lindfors}, E., {Nilsson}, K., {Barres de Almeida}, U., {et~al.} 2013,
  arXiv:1303.2102

\bibitem[{{Long} \& {Vachaspati}(2015)}]{2015JCAP...09..065L}
{Long}, A.~J., \& {Vachaspati}, T. 2015, \jcap, 9, 065

\bibitem[{{Neronov} {et~al.}(2010){Neronov}, {Semikoz}, {Kachelriess},
  {Ostapchenko}, \& {Elyiv}}]{2010ApJ...719L.130N}
{Neronov}, A., {Semikoz}, D., {Kachelriess}, M., {Ostapchenko}, S., \& {Elyiv},
  A. 2010, \apjl, 719, L130

\bibitem[{{Neronov} \& {Semikoz}(2009)}]{Nero-Semi:09}
{Neronov}, A., \& {Semikoz}, D.~V. 2009, \prd, 80, 123012

\bibitem[{{Neronov} \& {Vovk}(2010)}]{Nero-Vovk:10}
{Neronov}, A., \& {Vovk}, I. 2010, Science, 328, 73

\bibitem[{{Padovani} {et~al.}(2017){Padovani}, {Alexander}, {Assef}, {De
  Marco}, {Giommi}, {Hickox}, {Richards}, {Smol{\v c}i{\'c}}, {Hatziminaoglou},
  {Mainieri}, \& {Salvato}}]{Padovani+17}
{Padovani}, P., {Alexander}, D.~M., {Assef}, R.~J., {et~al.} 2017, \aapr, 25, 2

\bibitem[{{Pfrommer} {et~al.}(2012){Pfrommer}, {Chang}, \&
  {Broderick}}]{PaperIII}
{Pfrommer}, C., {Chang}, P., \& {Broderick}, A.~E. 2012, \apj, 752, 24

\bibitem[{{Puchwein} {et~al.}(2012){Puchwein}, {Pfrommer}, {Springel},
  {Broderick}, \& {Chang}}]{PaperIV}
{Puchwein}, E., {Pfrommer}, C., {Springel}, V., {Broderick}, A.~E., \& {Chang},
  P. 2012, \mnras, 423, 149

\bibitem[{{Pushkarev} {et~al.}(2009){Pushkarev}, {Kovalev}, {Lister}, \&
  {Savolainen}}]{Push_etal:09}
{Pushkarev}, A.~B., {Kovalev}, Y.~Y., {Lister}, M.~L., \& {Savolainen}, T.
  2009, \aap, 507, L33

\bibitem[{{Romoli} {et~al.}(2017){Romoli}, {Zacharias}, {Meyer}, {Ait
  Benkhali}, {Jacholkowska}, {Wierzcholska}, {Jankowsky}, {Lenain}, \& {for the
  H.~E.~S.~S.~Collaboration}}]{HESS_3c279_2015}
{Romoli}, C., {Zacharias}, M., {Meyer}, M., {et~al.} 2017, arXiv:1708.00882

\bibitem[{{Ryu} {et~al.}(2012){Ryu}, {Schleicher}, {Treumann}, {Tsagas}, \&
  {Widrow}}]{2012SSRv..166....1R}
{Ryu}, D., {Schleicher}, D.~R.~G., {Treumann}, R.~A., {Tsagas}, C.~G., \&
  {Widrow}, L.~M. 2012, \ssr, 166, 1

\bibitem[{{Schlickeiser} {et~al.}(2013){Schlickeiser}, {Krakau}, \&
  {Supsar}}]{Schlickeiser:2013}
{Schlickeiser}, R., {Krakau}, S., \& {Supsar}, M. 2013, \apj, 777, 49

\bibitem[{{Shalaby} {et~al.}(2017){Shalaby}, {Broderick}, {Chang}, {Pfrommer},
  {Lamberts}, \& {Puchwein}}]{Shalaby:resol}
{Shalaby}, M., {Broderick}, A.~E., {Chang}, P., {et~al.} 2017, \apj, 848, 81

\bibitem[{{Shalaby} {et~al.}(2018){Shalaby}, {Broderick}, {Chang}, {Pfrommer},
  {Lamberts}, \& {Puchwein}}]{Shalaby:inhomogeneity}
---. 2018, \apj, 859, 45

\bibitem[{{Stecker} {et~al.}(1992){Stecker}, {de Jager}, \&
  {Salamon}}]{Stec-deJa-Sala:92}
{Stecker}, F.~W., {de Jager}, O.~C., \& {Salamon}, M.~H. 1992, \apjl, 390, L49

\bibitem[{{Takahashi} {et~al.}(2012){Takahashi}, {Mori}, {Ichiki}, \&
  {Inoue}}]{Taka_etal:11}
{Takahashi}, K., {Mori}, M., {Ichiki}, K., \& {Inoue}, S. 2012, \apjl, 744, L7

\bibitem[{{Taylor} {et~al.}(2011){Taylor}, {Vovk}, \&
  {Neronov}}]{Tayl-Vovk-Nero:11}
{Taylor}, A.~M., {Vovk}, I., \& {Neronov}, A. 2011, \aap, 529, A144

\bibitem[{{Tiede} {et~al.}(2017{\natexlab{a}}){Tiede}, {Broderick}, {Shalaby},
  {Pfrommer}, {Puchwein}, {Chang}, \& {Lamberts}}]{BTII}
{Tiede}, P., {Broderick}, A.~E., {Shalaby}, M., {et~al.} 2017{\natexlab{a}},
  \apj, 850, 157

\bibitem[{{Tiede} {et~al.}(2017{\natexlab{b}}){Tiede}, {Broderick}, {Shalaby},
  {Pfrommer}, {Puchwein}, {Chang}, \& {Lamberts}}]{BTIII}
---. 2017{\natexlab{b}}, {arXiv:1702.02586}

\bibitem[{{Vafin} {et~al.}(2018){Vafin}, {Rafighi}, {Pohl}, \&
  {Niemiec}}]{Vafin18}
{Vafin}, S., {Rafighi}, I., {Pohl}, M., \& {Niemiec}, J. 2018, \apj, 857, 43

\bibitem[{{van Velzen} {et~al.}(2015){van Velzen}, {Falcke}, \&
  {K{\"o}rding}}]{dubbeltjes}
{van Velzen}, S., {Falcke}, H., \& {K{\"o}rding}, E. 2015, \mnras, 446, 2985

\bibitem[{{Vievering} {et~al.}(2015)}]{2015arXiv150807347J}
{Vievering}, J., {et~al.} 2015, arXiv:1508.07347

\bibitem[{{Vovk} {et~al.}(2012){Vovk}, {Taylor}, {Semikoz}, \&
  {Neronov}}]{Vovk+12}
{Vovk}, I., {Taylor}, A.~M., {Semikoz}, D., \& {Neronov}, A. 2012, \apjl, 747,
  L14

\bibitem[{{Wakely} \& {Horan}(2008)}]{TeVCat:2008}
{Wakely}, S.~P., \& {Horan}, D. 2008, International Cosmic Ray Conference, 3,
  1341

\bibitem[{{Widrow} {et~al.}(2012){Widrow}, {Ryu}, {Schleicher}, {Subramanian},
  {Tsagas}, \& {Treumann}}]{2012SSRv..166...37W}
{Widrow}, L.~M., {Ryu}, D., {Schleicher}, D.~R.~G., {et~al.} 2012, \ssr, 166,
  37

\bibitem[{{Zatsepin} \& {Kuz'min}(1966)}]{ZatsepinKuzmin1966}
{Zatsepin}, G.~T., \& {Kuz'min}, V.~A. 1966, Soviet Journal of Experimental and
  Theoretical Physics Letters, 4, 78

\bibitem[{{Zirakashvili} \& {Aharonian}(2007)}]{Zirakashvili07}
{Zirakashvili}, V.~N., \& {Aharonian}, F. 2007, \aap, 465, 695

\end{thebibliography}

\begin{appendix}

\section{Summary of SED Fit Parameters} \label{app:fits}
\begin{figure*}
  \begin{center}
    \includegraphics[width=0.7\textwidth]{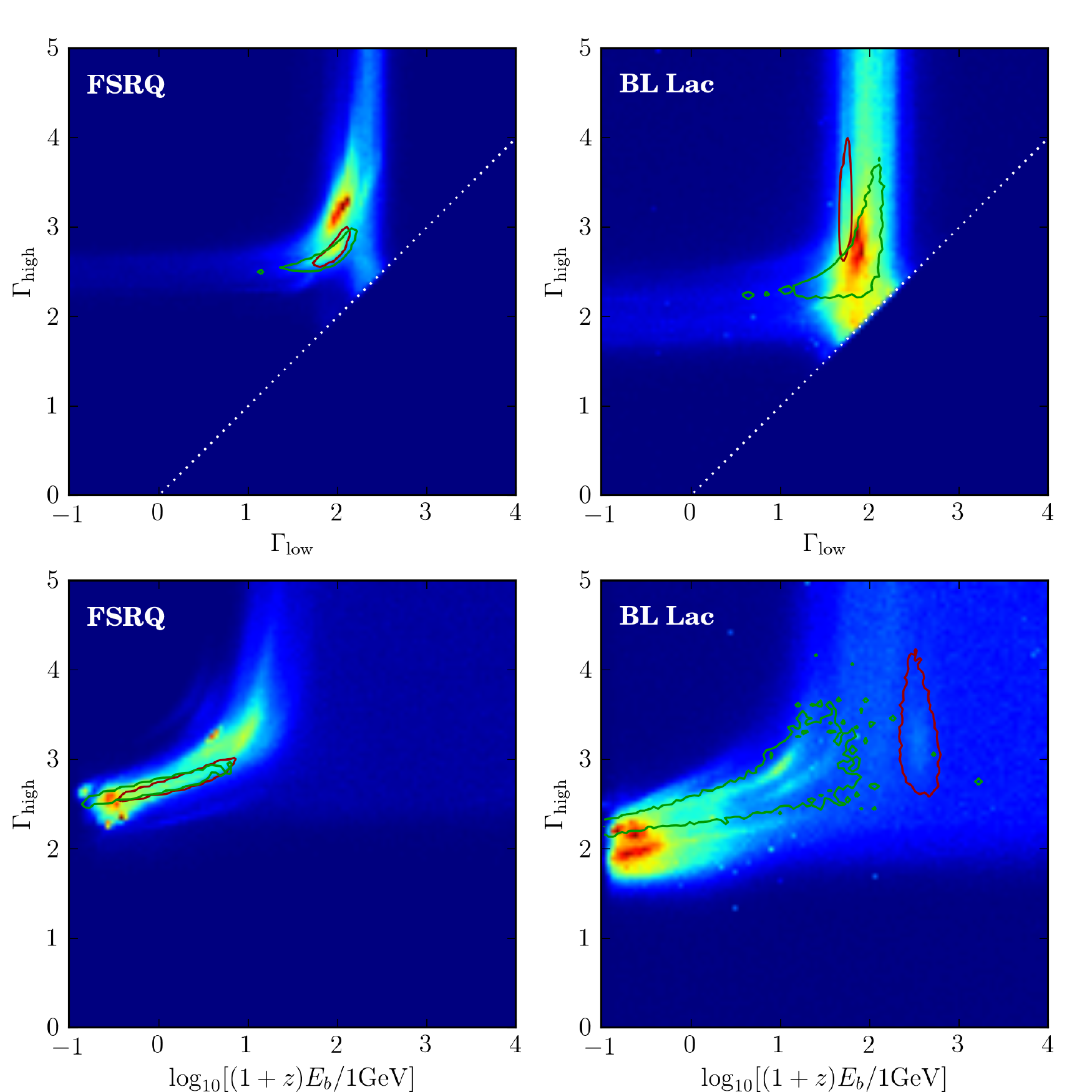}
  \end{center}
  \caption{Stacked joint distributions of the photon spectral indexes and rest-frame break energy of the SED fits to the FSRQs (left) and BL Lacs (right) that appear in the 3LAC and 3FHL with redshifts.  For reference, the 95\% confidence regions for the sources in Figure \ref{fig:05} are shown (FSRQs: 4C +21.35 (red), 3C 279 (green); BL Lacs: Mkn 421 (red), W Comae (green)).  Within the joint $\Gamma_{\rm low}$-$\Gamma_{\rm high}$ distributions, the $\Gamma_{\rm high}=\Gamma_{\rm low}$ line is indicated by the dotted white line.}\label{fig:fitsum}
\end{figure*}
We show a summary of some of the salient elements of the reconstructed SED parameters from our fits in Figure \ref{fig:fitsum}.  In particular, we show the joint distributions for the parameters that control the shape of the fitted SEDs: the high-energy photon spectral index, $\Gamma_{\rm high}\equiv\max(\Gamma_1,\Gamma_2)$, with the low-energy photon spectral index, $\Gamma_{\rm low}\equiv\min(\Gamma_1,\Gamma_2)$, and rest-frame break energy $(1+z)E_b$ (see Equation~\ref{eq:SEDfitfunc}).  The presence of individual islands in the stacked posterior distributions is a consequence of individual sources with well-constrained SEDs.  There does not appear to be a strong dependence on TeV luminosity.

The mean $\Gamma_{\rm high}$ depends on source class, with a value near 3.3 for FSRQs and 3.1 for BL Lacs.  We recover the typical TeV-GeV spectral break of unity found in \citet{2LAC}, here in the form of a generally positive $\Gamma_{\rm high}-\Gamma_{\rm low}$ near one; the mean break for FSRQs is 1.3 and for BL Lacs is 1.4.  That is, our estimates of $\Gamma_{\rm high}$ are statistically consistent with the spectral softening observed at a TeV in prior studies.

The typical $\Gamma_{\rm high}$ also depends weakly on the location of the spectral break; sources with higher breaks typically are marginally softer at high energies.  When $E_b>(1+z)10$~GeV, the fits exhibit a wide range of high-energy spectra, in part due to the weaker constraints on that portion of the spectrum.

For a substantial minority of sources, the SED is featureless, and thus a wide range of $E_b$ and $\Gamma_{\rm high,low}$ are permitted.  Note also that $\Gamma_{\rm high}$ asymptotes to roughly 2.5 for FSRQs and 2.3 for BL Lacs when $\Gamma_{\rm low}$ is very small; that is, $E L_E$ is rapidly rising at low energies.  Importantly, there are no fitted SEDs that exhibit pathologically hard high-energy behaviors.  This remains true even when the spectrum is rapidly rising at low energies (i.e., $\Gamma_{\rm low}<1$).

\section{Gamma-ray Mean Free Path and Optical Depth} \label{sec:Dpp}
Because \citet{Dominguez11} provides the optical depth to Earth from a given redshift at a given observed frequency in tabulated form, it is necessary to construct the local gamma-ray mean free path.  These are related via
\begin{equation}
  \tau_E(E_{\rm obs},z) = \int_0^z dz'\,\frac{dD_P}{dz}(z') \frac{1}{D_{\gamma\gamma}[E_{\rm obs} (1+z'),z']}
  \quad\text{where}\quad
  \frac{dD_P}{dz}(z) = \frac{1}{1+z} \frac{dD_C}{dz}(z),
\end{equation}
in which $D_P$ is the proper distance and $D_C$ is the standard comoving distance.  Therefore, in terms of the observed gamma-ray energy,
\begin{equation}
  D_{\gamma\gamma}[E_{\rm obs}(1+z),z]
  =
  \left(\frac{\partial \tau_E}{\partial z}\right)^{-1}
  \frac{dD_P}{dz}(z),
\end{equation}
where the partial derivative holds $E_{\rm obs}$ fixed.  To convert this to the local gamma-ray energy, we note that it is related to the observed gamma-ray energy via $E=E_{\rm obs}(1+z)$.

\section{FR II Redshift Distribution} \label{sec:FRIIzs}

\begin{figure*}
  \begin{center}
    \includegraphics[width=0.45\textwidth]{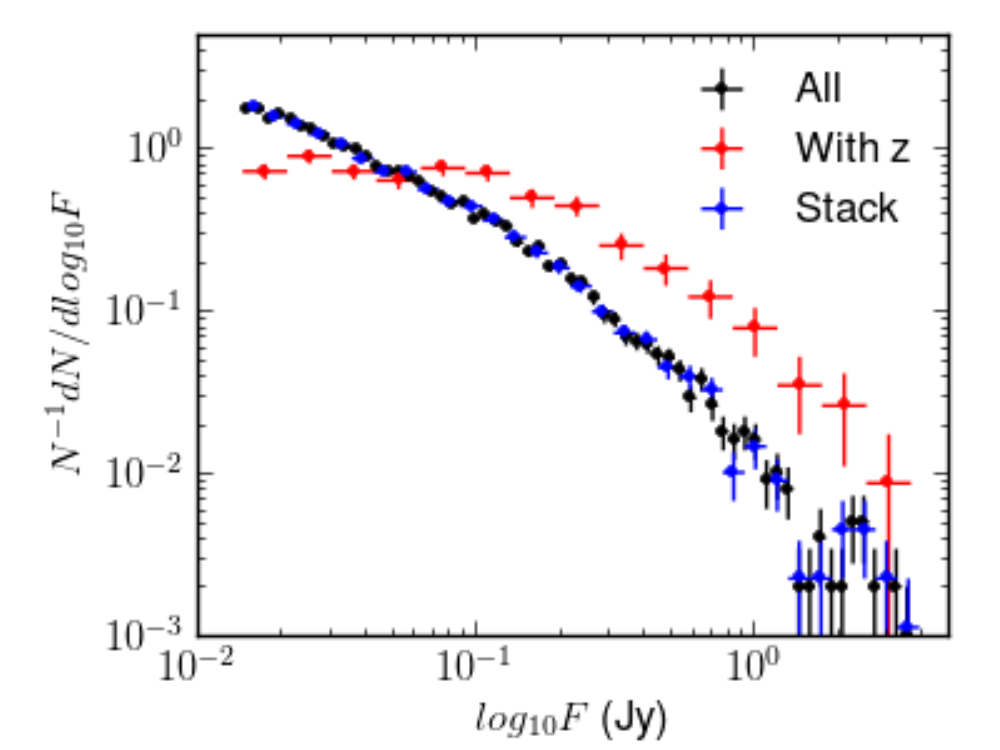}
    \includegraphics[width=0.45\textwidth]{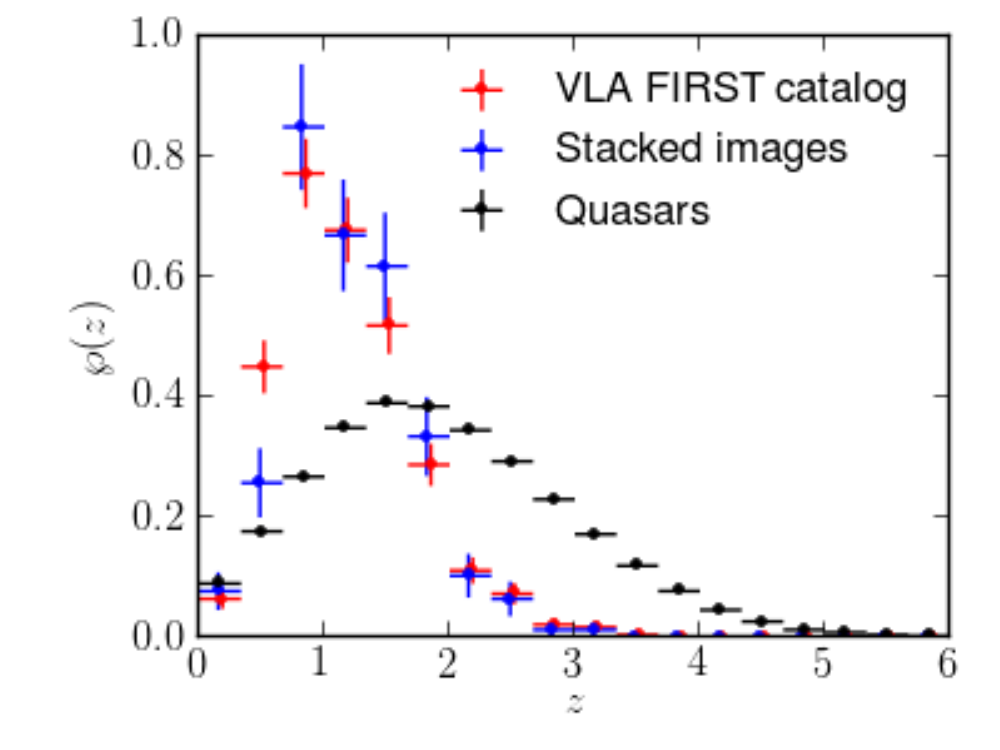}
  \end{center}
  \caption{Left: Flux distributions of the 24,973 double radio sources which meet the initial quality criteria (black), the 8,741 that are stacked in our FR II/FSRQ analysis (blue), and the subset of those with measured redshifts (red).
    Right: Redshift distribution of all sources with measured redshifts in the double-source VLA FIRST catalog (red) and of those used in the FR II object stacking analysis (blue).  
  \label{fig:06}}
\end{figure*}

Redshifts are measured for just 2.8\% of double radio sources that meet the initial quality control criteria listed in \cite{dubbeltjes}, the distribution of which is shown in Figure~\ref{fig:06}.  This fraction remains consistent when our additional quality criteria are applied.  However, the flux distributions of the sources with and without redshifts are clearly distinct, with the former typically being brighter than the latter (shown in Figure~\ref{fig:06}), suggesting that the
sources with redshifts are biased toward low $z$.

Therefore, we make
the pessimistic assumption that the redshift distribution of the radio
doubles is similar to that obtained from the quasar luminosity
function reported by \cite{Hopkins+07}, after applying the flux limits
stated in \cite{dubbeltjes}.  That is, we apply a bolometric
luminosity limit of
\begin{equation}
  L_{\rm min}(z) = 10^{3.57} 4\pi D_L^2(z) S_\nu (1.4{\rm GHz}) (1+z)^{-1.85},
\end{equation}
where the first term is a multiplicative correction between the radio
lobe and accretion disk luminosities, $S_\nu=12$~mJy is the 1.4~GHz
flux limit, and the final term is the band correction necessary to
obtain the rest-frame radio luminosity \citep{dubbeltjes}.
Note that we explicitly do not apply the FR II fraction correction reported in \citet{dubbeltjes}, as this is obtained from the subset of objects with redshifts; doing so would result in a larger number of lower-$z$ objects.
Therefore, the average redshift grows from 1.2 to 2.0, implying that the
radio doubles are systematically at higher redshifts.  The associated redshift distributions are shown in Figure~\ref{fig:06}.

\end{appendix}
\end{document}